\documentclass[%
 reprint,
amsmath,
amssymb,
aps,
floatfix,
]{revtex4-2}

\usepackage{graphicx}
\usepackage{dcolumn}
\usepackage{bm}
\usepackage{nccmath}
\usepackage{amsmath}
\usepackage{bm}
\usepackage{empheq}
\usepackage{dcolumn}
\usepackage{mathtools}
\usepackage{tensor}
\usepackage{bm}
\usepackage{subfigure}
\usepackage{comment}
\providecommand{\abs}[1]{\lvert#1\rvert} \providecommand{\norm}[1]{\lVert#1\rVert}

\begin{document}

\preprint{APS/123-QED}

\title{Charged spherically symmetric black holes in the Lyra geometry \\ and a preliminary investigation on the overcharging process}%

\author{Felipe Sobrero}
\email{felipesobrero@cbpf.br \\ felipesobrero@outlook.com}

\author{E. C. Valadão}%
 \email{eduvaladao98@cbpf.br \\ eduardovaladao98@gmail.com}
\affiliation{%
 Centro Brasileiro de Pesquisas Físicas (CBPF), Rio de Janeiro, CEP 22290-180, Brazil \\
}%

\begin{abstract}
This paper aims to investigate charged spherically symmetric static black holes in the Lyra geometry, in which a scale function naturally arises in the metric and affine structure of these type of manifolds. In particular, it is utilized the appropriate generalization of General Relativity, the recently proposed Lyra Scalar-Tensor Theory (LyST). The simplest generalization of Maxwell electrodynamics for Lyra manifolds is considered. It is presented an analytic solution for the line element of a Reissner–Nordström LyST generalization. It is shown that, due to the natural presence of a scale radius, it is possible to have three different extremal charges for positive or negative charge intervals. As a consequence, in natural units, the equality of the mass and charge defined on Lyra manifolds does not give rise to an extremal black hole, which allows the existence of solutions in which the charge is greater than the mass. An analysis with charged test particles indicates that a finite positive Lyra scale radius possibly allows for a violation of the weak cosmic censorship on Lyra manifolds, it is shown that an extremal black hole can be overcharged to the point that the emergence of a naked singularity becomes possible. The same behavior is observed for negative values of the Lyra radius if it's absolute value is greater than four times the black hole mass. Notably, this investigation also shows that an eternal black hole can exist for any charge increase if the Lyra scale radius is sufficiently close to some critical values. 

\end{abstract}

\maketitle

\section{Introduction}
\label{sec:1}

Observational evidence for black holes in different mass ranges has greatly increased in recent years. These include the direct detection of gravitational waves from binary mergers by the LIGO and Virgo interferometers \cite{Abbott_2019a,Abbott_2019b}, the first direct imaging of the M87* and Sgr A* shadows by the Event Horizon Telescope (EHT) collaboration \cite{Akiyama_2019a,Akiyama_2019b,Akiyama_2019c,Akiyama_2019d,Akiyama_2019e,Akiyama_2019f,Akiyama_2021a,Akiyama_2021b,Medeiros_2023} and the recently announced stochastic gravitational-wave background detected by the International Pulsar Timing Array (IPTA) \cite{Xu_2023,antoniadis2023second,Reardon_2023,Agazie_2023a}, which has been shown to be consistent with a population of binary supermassive black hole systems \cite{Agazie_2023b}.

From the theoretical perspective, black hole physics poses many fundamental questions about the nature of singularities \cite{Penrose_1965,Penrose_1973,Penrose_1993,Penrose_1999,Kerr2023rpn} and of the inconsistencies between Quantum Field Theory and General Relativity \cite{Hawking_2005}. Moreover, there has even been attempts to describe the dark sector of the standard cosmological model $\Lambda$CDM by black holes in different mass intervals \cite{Frampton_2010,Cappelluti_2022,Croker_2019,Farrah_2023}. As a result, the analysis of black hole solutions in different gravitational theories is of great importance, it can guide us in the construction of better models and assist in the physical interpretation of the growing amount of data of these astrophysical objects. 

These solutions, when compared to the ones of Einstein's theory of gravitation, allow us to improve our understanding of black holes and gravity itself. In special, due to the geometrical nature of spacetime, it is important to consider these physical objects in manifolds with different metric and affine structures. For this reason, this paper focus on a generalization of Riemannian geometry which was first presented by Lyra in 1951 \cite{Lyra_1951}. This metric-affine geometry introduces a scalar field, called Lyra scale function, which modifies vector lengths in a similar manner as the Weyl Integrable Spacetimes (WIST) \cite{Weyl_1922,Ross_1972,Novello_1983}.

Not long after, in 1957, Sen proposed what is considered the first theory of gravitation in a Lyra manifold \cite{Sen_1957}. Notably, the static cosmological solution to his vector-tensor theory accounts for the observed redshift of galactic spectral lines at the linear level \cite{Sen_1957}. Afterward, due to the intrinsic geometrical meaning of the Lyra scale function, Sen and Dunn proposed in 1971 a scalar-tensor theory on this geometry \cite{Sen_1971}. The resulting field equations are dynamically equivalent to a specific case of the Brans-Dicke theory in vacuum \cite{Sen_1971}.

However, Jeavons, McIntosh and Sen showed, in 1975, that the proposed equation of Sen and Dunn missed important contributions that were neglected from the variational principle utilized and, therefore, presented the corrected expression in \cite{Jeavons_1975}. Much later, in 2021, Cuzinatto, de Morais and Pimentel proposed that the use of an auxiliary vector field \textit{a priori} uncorrelated with the Lyra scale function as a mean to obtain a scalar-tensor theory is unjustified \cite{Cuzinatto_2021}, since a relation between these two fields is a natural  consequence of the variational principle utilized by \cite{Jeavons_1975}.

Therefore, the authors of \cite{Cuzinatto_2021} presented the so-called Lyra Scalar-Tensor Theory (LyST), which utilizes the metric tensor and the Lyra scale function as the fundamental fields. The LyST, due to its simplicity and geometrical meaning, is to be regarded as the proper generalization of General Relativity on Lyra manifolds \cite{Cuzinatto_2021}. They further presented a vacuum spherically symmetric solution, such that the metric tensor components act as the de Sitter or anti-de Sitter ones in the limit of large distances from the source \cite{Cuzinatto_2021}.

The LyST is the result of many years of scientific research, dating back to the works of Weyl \cite{Weyl_1918} and Dirac \cite{Dirac_1973}, in constructing a gravitational theory that is also covariant by means of a scale (or gauge) transformation \cite{Maeder_1978}. As a matter of fact, the Lyra geometry is the appropriate starting point to construct a scale covariant field theory. As a consequence of the structure added by the scale function and due to the natural appearence of a nonsymmetric connection in torsionless and metric compatible manifolds, the Lyra geometry is an important approach to the study of massless fields \cite{Maeder_1978} and spinorial ones \cite{Casana_2006}. Thereby, rendering it useful for the study of quantum theories of gravity \cite{Casana_2006}.

In view of these relevant aspects, this work utilizes a generalization of Maxwell's electromagnetism for Lyra manifolds as to study charged spherically symmetric black hole solutions in this geometry. We present in section \ref{sec2}, based on the more complete studies of \cite{Lyra_1951, Sen_1971, Cuzinatto_2021, Scheibe_1952}, a brief review of the important concepts and structures of the Lyra geometry. In section \ref{sec3}, it is shown the procedure to find the LyST field equations from an appropriate variational principle \cite{Cuzinatto_2021}. The Maxwell-Lyra energy-momentum tensor and the Reissner–Nordström LyST generalization line element are presented in section \ref{sec4}, along with a analysis of the horizons, singularities and geodesics of this new spacetime. In section \ref{sec5} we attempt to overcharge our black hole solution with charged particles and without considering backreaction. Finally, section \ref{sec6} features our final remarks and prospects for future research in Lyra manifolds.

\section{\label{sec2}Lyra geometry}

A $\mathcal{C^{\infty}}$ differentiable Lyra manifold $\mathcal{M_{L}}$ is a real set of dimension $n$ equipped with a maximal atlas $\mathcal{A_{L}} := \{(\mathcal{O}_{k}; \mathcal{X}_{k}, \Phi_{k})\}$ \cite{Sen_1971}. A Lyra reference system \cite{Lyra_1951} is defined as a triad $(\mathcal{O}_{k}; \mathcal{X}_{k}, \Phi_{k}) \in \mathcal{A_{L}}$, such that the $\mathcal{C^{\infty}}$ differentiable map $\mathcal{X}_{k}\text{:} \ \mathcal{O}_{k} \rightarrow \mathbb{R}^{n}$ is a chart over the open subset $\mathcal{O}_{k} \subset \mathcal{M_{L}}$ and $\Phi_{k}\text{:} \ \mathcal{O}_{k} \rightarrow \mathbb{R}^{*}$ is a $\mathcal{C^{\infty}}$ scale map on this subset of the manifold. Furthermore, if $p \in \mathcal{O}_{k}$ is a point of the manifold, $x(p) := \mathcal{X}_{k} \circ p$ are its coordinates on the chart $\mathcal{X}_{k}$, therefore, the scale map at $p$ can be written in terms of these coordinates as $\Phi_{k} \circ p = \Phi_{k} \circ \mathcal{X}^{-1}_{k} \circ x(p) \coloneq \phi(x(p))$. Consequently, the map $\phi\text{:} \ \mathbb{R}^{n} \rightarrow \mathbb{R}^{*}$, which is called Lyra scale function \cite{Lyra_1951}, is defined as $\phi \coloneq \Phi_{k} \circ \mathcal{X}^{-1}_{k}$ \cite{Cuzinatto_2021}. 

If $\mathcal{F}$ is the set of all $\mathcal{C^{\infty}}$ functions $f\text{:} \ \mathcal{O}_{k} \rightarrow \mathbb{R}$, a tangent vector at $p$ is defined as the linear map $\mathbf{v}\text{:} \ \mathcal{F} \rightarrow \mathbb{R}$ which respects the Leibniz product rule. As a result, the set $\mathcal{T}_{p}$ of all tangent vectors at $p$ form a vector space if we assume that the scalar multiplication and addition law are satisfied by the elements of this space. A canonical basis for $\mathcal{T}_{p}$, which obeys all these rules, can be constructed by defining:
\begin{equation}
\label{eq1}
\mathbf{e}_{\mu}f \coloneq \frac{1}{\phi(x)} \frac{\partial (f \circ \mathcal{X}^{-1}_{k})}{\partial x^{\mu}}\Biggr|_{\substack{x(p)}},
\end{equation}
in which $x^{\mu}$ are the coordinates in the chart $\mathcal{X}_{k}$ \cite{Sen_1971}. Thus, the fundamental difference between this geometry and Riemannian ones is in the definition of this new basis with noncommutative elements \cite{Cuzinatto_2021}.

A tangent vector can then be written as $\mathbf{v} = v^{\mu} \mathbf{e}_{\mu}$, with $v^{\mu}$ being its components. Therefore, the basis elements of $\mathbf{v}$ under a transformation of Lyra reference systems, from $(\mathcal{O}; \mathcal{X}, \Phi)$ to $(\bar{\mathcal{O}}; \bar{\mathcal{X}}, \bar{\Phi})$, change accordingly to:
\begin{equation}
\label{eq2}
\bar{\mathbf{e}}_{\mu} = \frac{\phi (x)}{\bar{\phi} (\bar{x})} \frac{\partial x^{\nu}}{\partial \bar{x}^{\mu}} \mathbf{e}_{\nu},
\end{equation}
and, consequently, the components must transform as: 
\begin{equation}
\label{eq3}
\bar{v}^{\mu} = \frac{\bar{\phi} (\bar{x})}{\phi (x)} \frac{\partial \bar{x}^{\mu}}{\partial x^{\nu}} v^{\nu}.
\end{equation}
Hence, a change between different Lyra reference systems is simultaneously a coordinate and a scale transformation. Moreover, we can define a tangent vector to a smooth curve $\gamma\text{:} \ \lambda \in \mathbb{R} \rightarrow \mathcal{M_{L}}$ at $p$ as the directional derivative:

\begin{equation}
\label{eq4}
\mathbf{v}_{\gamma}f \coloneq \frac{d (f \circ \gamma)}{d \lambda}\Biggr|_{\substack{\lambda_{p}}} = \frac{d x^{\mu}}{d \lambda} \partial_{\mu} (f \circ \mathcal{X}^{-1}_{k})\Biggr|_{\substack{x(p)}},
\end{equation}
therefore, the components of the tangent vector to the curve $\gamma(\lambda)$ in the basis \eqref{eq1} of chart $\mathcal{X}_{k}$ are defined as:
\begin{equation}
\label{eq5}
v^{\mu} = \phi(x) \frac{d x^{\mu}}{d \lambda}.
\end{equation} 

As it is in Riemannian geometry, a tangent vector space $\mathcal{T}_{p}$ allow us to define linear maps $\bm{\omega}\text{:} \ \mathcal{T}_{p} \rightarrow \mathbb{R}$ called dual vectors, which defines the dual vector space $\mathcal{T}^{*}_{p}$. The relation \eqref{eq4} makes possible to define natural dual vectors as:
\begin{equation}
\label{eq6}   
\mathbf{d}f = \frac{1}{\phi (x)} \partial_{\mu} (f \circ \mathcal{X}^{-1}_{k})\Biggr|_{\substack{x(p)}}\bm{\theta}^{\mu},
\end{equation}
since $\mathbf{d}f \circ \mathbf{v} = df_{\mu} v^{\mu}$, due to the orthonormality condition $\bm{\theta}^{\mu} \circ \mathbf{e}_{\nu} = \tensor{\delta}{^\mu_\nu}$, is defined to be equal to expression \eqref{eq4}. A natural basis for $\mathcal{T}^{*}_{p}$ can then, using the orthonormality condition, be defined as \cite{Cuzinatto_2021}:
\begin{equation}
\label{eq7}
\bm{\theta}^{\mu} = \phi(x) \mathbf{d}x^{\mu}.
\end{equation}
Furthermore, by making $\bm{\omega} \circ \bar{\mathbf{e}}_{\mu}$, it is possible to find that the components $\omega_{\mu}$ of a dual vector transform accordingly to expression \eqref{eq2}, and, as a consequence, its basis elements $\bm{\theta}^{\mu}$ transform as in relation \eqref{eq3}.

With the local vector space and its dual defined, it is possible to define multilinear maps that take $k$ elements of $\mathbf{T}^{*}_{p}$ and $l$ vectors of $\mathbf{T}_{p}$ to the real numbers as objects called Lyra tensors and defined as:
\begin{equation}
\label{eq8}
T = T^{\mu_{1}...\mu_{k}}_{\ \ \ \ \ \ \ \ \ \nu_{1}...\nu_{l}} \bm{e}_{\mu_{1}} \otimes ... \otimes \bm{e}_{\mu_{k}} \otimes \bm{\theta}^{\nu_{1}} \otimes ... \otimes \bm{\theta}^{\nu_{l}},
\end{equation}
so that the sum, scalar product, tensorial product and contraction of these objects are defined as in Riemannian geometry. The solely difference, is that under a Lyra reference system change, tensors transform as \cite{Cuzinatto_2021}:
\begin{equation}
\label{eq9}
\hspace*{-0.1cm}
\bar{T}^{\mu_{1}...\mu_{k}}_{\ \ \ \ \ \ \ \ \nu_{1}...\nu_{l}}=\bigg(\frac{\bar{\phi} (\bar{x})}{\phi (x)}\bigg)^{k-l} \frac{\partial \bar{x}^{\mu_{1}}}{\partial x^{\eta_{1}}} ... \frac{\partial x^{\xi_{l}}}{\partial \bar{x}^{\nu_{l}}} T^{\eta_{1}...\eta_{k}}_{\ \ \ \ \ \ \ \xi_{1}...\xi_{l}}.
\end{equation}

\subsection{The metric structure} \label{subsec2.1}

For the concept of a manifold to be associated with the physical ideia of spacetime, it is necessary an additional structure that allows the definition of spatio-temporal lengths. The object that adds this causal structure definition is a smooth bilinear symmetric nondegenerate map $\mathbf{g}\text{:} \ \mathcal{T}_{p} \times \mathcal{T}_{p} \rightarrow \mathbb{R}$ named metric tensor. It is a family of inner products, so that at each point of $\mathcal{M_{L}}$ the inner product of two vectors, $\mathbf{v} = v^{\mu} \mathbf{e}_{\mu}$ and $\mathbf{u} = u^{\nu} \mathbf{e}_{\nu}$, is given by:
\begin{equation}
\label{eq10}
\mathbf{g}(\mathbf{v}, \mathbf{u}) = \tensor{g}{_\mu_\nu}v^{\mu}u^{\nu},
\end{equation}
since $\tensor{g}{_\mu_\nu} \coloneq \mathbf{g}(\mathbf{e}_{\mu}, \mathbf{e}_{\nu})$. This structure further adds the canonical identification of the $\mathcal{T}^{*}_{p}$ and $\mathcal{T}_{p}$ spaces by defining that $v_{\mu} \coloneq \mathbf{g}(\mathbf{v}, \mathbf{e}_{\mu}) = \tensor{g}{_\mu_\nu} v^{\nu}$. 

Moreover, as it is in pseudo-Riemannian geometry, the inner product of vectors is not necessarily positive definite due to the Lorentzian signature and the inverse metric is defined by $\tensor{g}{^\mu^\alpha} \tensor{g}{_\alpha_\nu} = \tensor{\delta}{^\mu_\nu}$. However, if we define vector lengths as $\norm{\mathbf{v}}^{2} \coloneq \mathbf{g}(\mathbf{v}, \mathbf{v})$, it is possible to see from \eqref{eq5} that:
\begin{equation}
\label{eq11}
\norm{\mathbf{v}} = \abs{\phi} \sqrt{\tensor{g}{_\mu_\nu} \frac{d x^{\mu}}{d \lambda} \frac{d x^{\nu}}{d \lambda}}.
\end{equation}
Thus, the Lyra scale function alters tangent vector lengths when compared to the definition in Riemannian manifolds. It can further be showed to add no second clock effects \cite{Romero_2019}, allowing normalizations like $v^{\mu} v_{\mu} = c^{2}$.

Consequently, the length of a curve $\gamma(\lambda)$, with tangent vector $\mathbf{v}$, from $\lambda_{1}$ to $\lambda_{2}$, is given by:
\begin{equation}
\label{eq12}
s = \int_{\lambda_{1}}^{\lambda_{1}} \abs{\phi} \sqrt{\tensor{g}{_\mu_\nu} \frac{d x^{\mu}}{d \lambda} \frac{d x^{\nu}}{d \lambda}} d\lambda,
\end{equation}
so that the line element, which is invariant under Lyra reference system transformations, is defined as:
\begin{equation}
\label{eq13} 
ds^{2} = \phi^{2} \tensor{g}{_\mu_\nu} dx^{\mu} dx^{\nu}.
\end{equation} 
If expression \eqref{eq12} is stationary for fixed extreme points in the configuration space, by doing $\delta s = 0$ and considering $\lambda$ an affine parameter we obtain the Lyra geodesic equation:
\begin{empheq}{align}
\label{eq14}
  &\frac{d^{2}x^{\gamma}}{ds^{2}} + \bigg\{\genfrac{}{}{0pt}{0}{\gamma}{\mu \nu}\bigg\}\frac{dx^{\mu}}{ds}\frac{dx^{\nu}}{ds} + \\
  &\frac{1}{\phi} (\tensor{\delta}{^\gamma_\nu} \partial_{\mu} \phi + \tensor{\delta}{^\gamma_\mu} \partial_{\nu} \phi - \tensor{g}{_\mu_\nu} \partial^{\gamma} \phi) \frac{dx^{\mu}}{ds}\frac{dx^{\nu}}{ds} = 0, \nonumber
\end{empheq}
which adds new terms that differ it from the metric geodesic equation of Riemannian manifolds.

Furthermore, since under a change of Lyra reference systems the determinant of the metric $g$ for $n$ dimensions transform as \cite{Cuzinatto_2021}:
\begin{equation}
\label{eq15}    
\bar{\phi}^{2n} (\bar{x}) \Bar{g} (\bar{x}) = \abs{J}^{2} \phi^{2n} (x) g(x),
\end{equation}
which is found by utilizing the transformation rule \eqref{eq9} and such that $\abs{J}$ is the Jacobian determinant of the transformation, the volume element in Lyra manifolds is modified and it is given by:
\begin{equation}
\label{eq16}
dV = \phi^{n} \sqrt{\abs{g}} d^{n}x,
\end{equation}
so that it is properly defined as a Lyra $n$-form.

\subsection{The affine structure} \label{subsec2.2}

To properly do physics on Lyra manifolds it is necessary to include another structure that connects the vector spaces of different points of $\mathcal{M_{L}}$. This structure is defined by the map $\bm{\nabla}\text{:} \ \mathcal{T}_{p} \times \mathcal{T}_{p} \rightarrow \mathcal{T}_{p}$. It takes two vectors, for example, $\mathbf{u}$ and $\mathbf{v}$, to the object $\bm{\nabla}_{\mathbf{u}} \mathbf{v}$. It is further required to be linear, that is, it must satisfy \cite{Cuzinatto_2021}:
\begin{equation}
\label{eq17}
\begin{cases}
\bm{\nabla}_{\mathbf{u}}(\mathbf{v} + \mathbf{w}) = \bm{\nabla}_{\mathbf{u}}\mathbf{v} + \bm{\nabla}_{\mathbf{u}}\mathbf{w}  \\
\bm{\nabla}_{\mathbf{u + v}}\mathbf{w} = \bm{\nabla}_{\mathbf{u}}\mathbf{w} + \bm{\nabla}_{\mathbf{v}}\mathbf{w}
\end{cases},
\end{equation}
for $\mathbf{w} \in \mathcal{T}_{p}$. If $f \in \mathcal{F}$, this linear object must also satisfy the properties:
\begin{equation}
\label{eq18}
\hspace*{-0.15cm}
\begin{cases}
\bm{\nabla}_{f\mathbf{u}}\mathbf{v} = f\bm{\nabla}_{\mathbf{u}}\mathbf{v} \\
\bm{\nabla}_{\mathbf{u}}(f\mathbf{v}) = (\mathbf{u}f)\mathbf{v} + f\bm{\nabla}_{\mathbf{u}}\mathbf{v}
\end{cases}.
\end{equation}
Therefore, by utilizing the canonical Lyra basis \eqref{eq1}, these properties lead us to \cite{Cuzinatto_2021}:
\begin{equation}
\label{eq19}
\bm{\nabla}_{\mathbf{u}}\mathbf{v} = u^{\nu} \big(\nabla_{\nu}v^{\alpha}\big) \mathbf{e}_{\alpha} = u^{\nu} \bigg(\frac{1}{\phi}\partial_{\nu}v^{\alpha} + \Gamma^{\alpha}_{\mu\nu}v^{\mu}\bigg) \mathbf{e}_{\alpha},
\end{equation}
in which $(\nabla_{\nu}v^{\alpha})\mathbf{e}_{\alpha} \coloneq \bm{\nabla}_{\mathbf{e}_{\nu}}\mathbf{v}$ defines the Lyra covariant derivative and $\Gamma^{\alpha}_{\mu\nu} \mathbf{e}_{\alpha} \coloneq \bm{\nabla}_{\mathbf{e}_{\nu}}\mathbf{e}_{\mu}$ its affine connection. 

It can be seen from \eqref{eq19} that the covariant derivative of a scalar function $f$ can naturally be defined as $\nabla_{\mu}f = \phi^{-1} \partial_{\mu} f$. Using this property on the covariant derivative of a vector contracted with a dual vector it is possible to find that this map when applied to dual vectors is given by: 
\begin{equation}
\label{eq20}
\nabla_{\mu} \omega_{\nu} = \frac{1}{\phi}\partial_{\mu} \omega_{\nu} - \Gamma^{\alpha}_{\nu\mu} \omega_{\alpha}.
\end{equation}
Therefore, the covariant derivative of a general tensor is almost similar to the Riemannian expression, the exception being the factor $\phi^{-1}$ multiplying the partial derivative and the presence of a different affine connection. 

The inclusion of an affine structure to $\mathcal{M_{L}}$ means that there is now a notion of parallelism between vectors on different points. Since tangent vectors are associated with directional derivatives on curves, a constant change of a vector $\mathbf{v}$ along a curve represented by $\mathbf{u}$ is given by $\bm{\nabla}_{\mathbf{u}} \bm{v} = \bm{0}$. It is then said that $\mathbf{v}$ is parallel transported in the direction of $\mathbf{u}$. Thus, $\bm{\nabla}_{\mathbf{v}} \mathbf{v} = \bm{0}$, which can be written as:
\begin{equation}
\label{eq21}
\frac{d^{2}x^{\alpha}}{d\lambda^{2}} + \big(\phi \Gamma^{\alpha}_{\mu\nu} + \tensor{\delta}{^\alpha_\mu} \nabla_{\nu}\phi \big) \frac{dx^{\mu}}{d\lambda}\frac{dx^{\nu}}{d\lambda} = 0,
\end{equation}
means that we are parallel transporting a vector along its own curve. For this reason, the curves whose tangent vectors satisfy this relation are then called autoparallel curves.

With this structure defined, it is straightforward to obtain other important geometrical entities. Since curvature and torsion produces non-commutativity of successive covariant differentiations of a tensor, using the Lyra covariant derivative defined by \eqref{eq19} allow us to obtain the Riemann tensor on Lyra manifolds:
\begin{equation}
\label{eq22}
\tensor{R}{^\rho_\mu_\gamma_\alpha} = \frac{2}{\phi^{2}}\partial_{[\gamma} \big(\phi \Gamma^{\rho}_{|\mu|\alpha]}\big) + 2\Gamma^{\rho}_{\sigma[\gamma} \Gamma^{\sigma}_{|\mu|\alpha]},
\end{equation}
which is antisymmetric in its last two indices. It is further possible to define the Lyra torsion tensor as:
\begin{equation}
\label{eq23}
\tensor{\tau}{^\rho_\gamma_\alpha} = \Gamma^{\rho}_{\alpha\gamma} - \Gamma^{\rho}_{\gamma\alpha} + \frac{1}{\phi} \big(\tensor{\delta}{^\rho_\alpha} \nabla_{\gamma} \phi -  \tensor{\delta}{^\rho_\gamma} \nabla_{\alpha} \phi\big),
\end{equation}
in which the new term is equal to the basis non-commutativity tensor \cite{Cuzinatto_2021}. Moreover, since the non-metricity tensor $\tensor{Q}{_\alpha_\mu_\nu} \coloneq - \nabla_{\alpha} \tensor{g}{_\mu_\nu}$ relates the metric and affine structure, doing $-\tensor{Q}{_\alpha_\mu_\nu} + \tensor{Q}{_\mu_\nu_\alpha} + \tensor{Q}{_\nu_\alpha_\mu}$ yields:
\begin{equation}
\label{eq24}
\Gamma^{\gamma}_{\mu\nu} = \frac{1}{\phi} \bigg\{\genfrac{}{}{0pt}{0}{\gamma}{\mu \nu}\bigg\} + \frac{1}{\phi}\big(\tensor{\delta}{^\gamma_\nu} \nabla_{\mu} \phi - \tensor{g}{_\mu_\nu}\nabla^{\gamma} \phi \big) + \tensor{N}{^\gamma_\mu_\nu}, 
\end{equation}
in which $\tensor{N}{^\gamma_\mu_\nu} = \tensor{K}{^\gamma_\mu_\nu} + \tensor{L}{^\gamma_\mu_\nu}$ is the Lyra distortion tensor, such that $\tensor{g}{^\gamma^\alpha} (\tensor{\xi}{_\mu_\alpha_\nu} + \tensor{\xi}{_\nu_\alpha_\mu} - \tensor{\xi}{_\alpha_\mu_\nu})/2$ is the contorsion tensor $\tensor{K}{^\gamma_\mu_\nu}$ if $\tensor{\xi}{_\alpha_\mu_\nu} = \tensor{\tau}{_\alpha_\mu_\nu}$ and $\tensor{L}{^\gamma_\mu_\nu}$ if $\tensor{\xi}{_\alpha_\mu_\nu} = \tensor{Q}{_\alpha_\mu_\nu}$.

\section{Lyra Scalar-Tensor Theory - LyST} \label{sec3}

To construct a generalization of General Relativity on Lyra manifolds from a metric variational principle it is first required that geodesics are autoparallel curves. By using \eqref{eq14}, \eqref{eq21} and \eqref{eq24} this imposition results in the condition $L_{\gamma(\mu\nu)} = \tau_{(\mu\nu)\gamma}$. It is also necessary a well defined divergence theorem version for Lyra manifolds to properly deal with surface terms, this can be simply done if $Q_{\gamma\mu\nu} = - 2 \tau_{(\mu\nu)\gamma}$. As a consequence of these two conditions, we have that $Q_{(\gamma\mu\nu)} = 0$. However, since it is required a version of Einstein's theory in the Lyra geometry, it is necessary to consider only torsionless and metric compatible manifolds, which automatically satisfy the above conditions \cite{Cuzinatto_2021}.

These considerations lead to the vanishing of the distortion tensor in the Lyra linear connection \eqref{eq24}. It further allow us to separate the Riemann tensor \eqref{eq22} in a Riemannian part and in derivatives of the Lyra scale function, so that the nonsymmetric Ricci tensor is written as:

\begin{eqnarray}\label{eq25}
  \tensor{R}{_\mu_\nu} &=& \frac{1}{\phi^{2}} \tensor{\mathcal{R}}{_\mu_\nu} \ - \ \frac{2}{\phi} \nabla_{\nu} \nabla_{\mu} \phi \ - \ \frac{1}{\phi} \tensor{g}{_\mu_\nu} \Box \phi \nonumber \\
  && + \frac{3}{\phi^{2}} \tensor{g}{_\mu_\nu} {\nabla^{\rho} \phi} {\nabla_{\rho} \phi}, 
\end{eqnarray}

\noindent in which $\tensor{\mathcal{R}}{_\mu_\nu}$ is defined as the Ricci tensor of Riemannian geometry and such that $\Box \coloneq \nabla^\rho \nabla_{\rho}$ is the d'Alembertian generalization on Lyra manifolds. Therefore, the Ricci scalar in a four-dimensional manifold is given by:
\begin{equation}
\label{eq26}
R = \frac{1}{\phi^{2}} \mathcal{R} \ - \ \frac{6}{\phi} \Box \phi \ + \ \frac{12}{\phi^{2}} {\nabla^{\rho} \phi} {\nabla_{\rho} \phi},
\end{equation}
in which $\mathcal{R} \coloneq \tensor{g}{^\mu^\nu} \tensor{\mathcal{R}}{_\mu_\nu}$ has the same definition as the Ricci scalar of Riemannian manifolds.

By using expression \eqref{eq26}, and considering the volume element \eqref{eq16}, is then possible to write the four-dimensional action for the Lyra Scalar-Tensor Theory as \cite{Cuzinatto_2021}:
\begin{equation}
\label{eq27}
S_{LyST} = \frac{1}{c} \frac{1}{2\kappa} \int_{\Omega} R \phi^{4} \sqrt{\abs{g}} d^{4}x + S_{m}\big(\psi; \phi, \tensor{g}{_\mu_\nu}\big),
\end{equation}
in which the matter action $S_{m}$ depends on the matter fields $\psi$ and on the geometrical fields, $\tensor{g}{_\mu_\nu}$ and $\phi$, and its derivatives $\partial_{\alpha} \tensor{g}{_\mu_\nu}, \partial_{\gamma} \partial_{\alpha} \tensor{g}{_\mu_\nu}$ and $\nabla_{\mu} \phi = \phi^{-1} \partial_{\mu} \phi$. The LyST field equations are then obtained by the variation of this action with respect to the metric tensor, if the surface terms are ignored, this action yields \cite{Cuzinatto_2021}:
\begin{eqnarray}
\label{eq28}
  \frac{1}{\phi^{2}} \tensor{\mathcal{G}}{_\mu_\nu} \ - \ \frac{2}{\phi} \nabla_{(\mu} \nabla_{\nu)} \phi \ + \ \frac{2}{\phi} \tensor{g}{_\mu_\nu} \Box \phi && \\
   -  \frac{3}{\phi^{2}} \tensor{g}{_\mu_\nu} {\nabla^{\rho} \phi} {\nabla_{\rho} \phi} &=& \kappa \tensor{T}{_\mu_\nu}, \nonumber
\end{eqnarray}
such that $\tensor{\mathcal{G}}{_\mu_\nu} \coloneq \tensor{\mathcal{R}}{_\mu_\nu} - \frac{1}{2} \mathcal{R} \tensor{g}{_\mu_\nu}$ and the energy-momentum tensor is defined as:
\begin{equation}
\label{eq29}
\tensor{T}{_\mu_\nu} \coloneq \frac{-2}{\sqrt{\abs{g}}} \frac{\delta \big(\mathcal{L}_{m} \sqrt{\abs{g}} \big)}{\delta \tensor{g}{^\mu^\nu}}.
\end{equation}

The left-hand side of equation \eqref{eq28} is just the symmetric part of the Einstein tensor $\tensor{G}{_\mu_\nu}$ generalized on Lyra manifolds, as a matter of fact, Riemannian geometry and General Relativity are obtained in the special case of $\phi = 1$. As a consequence, it is justified to set $\kappa = 8 \pi G c^{-4}$. Additionally, if we vary the action \eqref{eq27} with respect to $\phi$ we obtain that $R = \kappa M$, in which:
\begin{equation}
\label{eq30}    
M = -4\mathcal{L}_{m} - \phi \bigg( \frac{\partial \mathcal{L}_{m}}{\partial \phi} - \nabla_{\mu} \frac{\partial \mathcal{L}_{m}}{\partial \nabla_{\mu} \phi} \bigg) + \frac{\partial \mathcal{L}_{m}}{\partial \nabla_{\mu} \phi} \nabla_{\mu} \phi.
\end{equation}
Taking the trace of the field equations \eqref{eq28} we find that $R = - \kappa T$, so that the second equation is given by $M = - T$. 

\section{Charged spherically symmetric solution} \label{sec4} 

One of the purposes of this paper is to find an analytic charged spherically symmetric solution to the LyST equations \eqref{eq28}. As a consequence, it is first necessary to generalize Maxwell's electromagnetism to Lyra manifolds. By using a minimal coupling prescription, the Maxwell-Lyra energy-momentum tensor is defined as:
\begin{equation}
\label{eq32}
\tensor{T}{_\mu_\nu} = \frac{1}{\mu_{0}} \bigg(\tensor{g}{^\alpha^\gamma} \tensor{F}{_\mu_\alpha} \tensor{F}{_\gamma_\nu} + \frac{1}{4} \tensor{g}{_\mu_\nu} \tensor{F}{_\alpha_\gamma} \tensor{F}{^\alpha^\gamma} \bigg),
\end{equation}
in which the generalized Faraday tensor is now defined with Lyra covariant derivatives:
\begin{equation}
\label{eq33}
\tensor{F}{_\mu_\nu} = 2\nabla_{[\mu} A_{\nu]} = 2\phi^{-1} \big(\partial_{[\mu} A_{\nu]} + A_{\alpha} \tensor{\delta}{^\alpha_{[\mu}} \nabla_{\nu]} \phi \big),
\end{equation}
such that the new term emerges from the effective torsion $2 \Gamma^{\alpha}_{[\nu\mu]}$. Consequently, the first Maxwell-Lyra equation in the absence of matter fields is simply given by:
\begin{equation}  
\label{eq34}    
\nabla_{\mu} \tensor{F}{^\mu^\nu} = 0,
\end{equation}
with the covariant derivative defined on Lyra manifolds.

On the geometrical part, it is assumed that the isometry group of the spatial part of the metric is the SO(3) one. It is also considered a static metric, such that its radial and temporal part are described by the function $\alpha = \alpha(r)$ as:
\begin{equation}
\label{eq35}
\tensor{g}{_\mu_\nu} = \text{diag}\left( \alpha, - \alpha^{-1}, -r^2, -r^2 \sin^{2} \theta \right).
\end{equation}
As a result of the spherical symmetry, $\phi = \phi(t,r)$ as to keep the line element \eqref{eq13} invariant by spatial rotations. However, it is also assumed that $\phi = \phi(r)$ so that the metric is static. Therefore, the four-potential $A_{\mu}$ of a static electric charge can be written as:
\begin{equation}
\label{eq36}
A_{\mu} = (f,0,0,0),
\end{equation}
in which $f = f(r)$ is solution of the equation \eqref{eq34}.

With these assumptions, and considering the prime symbol as indicating the derivative of a function with respect to $r$, the first two LyST field equations from \eqref{eq28} with the energy-momentum tensor \eqref{eq32} are then simply given by:

\begin{eqnarray}
    \hspace{-0.7cm}
    \frac{1}{r^2} - \frac{1}{r^2 \alpha} + \frac{\alpha'}{r \alpha} + \frac{4 \phi'}{r \phi} + \frac{\alpha' \phi'}{\alpha \phi} -\frac{\phi'^2}{\phi^2} + \frac{2 \phi''}{\phi} &=& \zeta(r),\label{eq37} \\
    \frac{1}{r^2} - \frac{1}{r^2 \alpha} + \frac{\alpha'}{r \alpha} + \frac{4 \phi'}{r \phi} + \frac{\alpha' \phi'}{\alpha \phi} + \frac{3 \phi'^2}{\phi^2} &=& \zeta(r), \label{eq38} 
\end{eqnarray} 

\noindent in which the function $\zeta(r)$ corresponds to the electromagnetic part:
\begin{equation}
\label{eq39}  
\zeta(r) = -\frac{\kappa}{2 \mu_{0} \alpha \phi^{2}}(\phi f' + \phi' f)^{2}. 
\end{equation}
If we subtract eq. \eqref{eq38} from \eqref{eq37} it is possible to obtain a equation for the scale function:
\begin{equation}
\label{eq40}
\phi'' - \frac{2 \phi'^2}{\phi} = 0,
\end{equation}
such that its solution can be expressed as:
\begin{equation}
\label{eq41}
\phi(r) = \frac{r_{0}/r_{L}}{1 - r/r_{L}},
\end{equation}
in which $r_{L}, r_{0} \in \mathbb{R}$ are integration constants. From this expression, it can be seen that the field $\phi(r)$ naturally induces a characteristic scale $r_{L}$ on these type of manifolds, leading to the designation of this parameter as the Lyra scale radius.

Regarding the electric part, the first Maxwell-Lyra equation obtained from the definition \eqref{eq34} is given by:
\begin{equation}
\label{eq42}
f'' + \frac{2 f' \phi'}{\phi} + \frac{2 f'}{r} + \frac{2 f \phi'}{r \phi} + \frac{f \phi''}{\phi} = 0,
\end{equation}
which results in the static potential solution:
\begin{equation}
\label{eq43}
f(r) = \bigg( \frac{1}{4 \pi \epsilon_{0} c} \frac{Q_L}{r} + c_{1} \bigg) \bigg( 1 - \frac{r}{r_L} \bigg),
\end{equation}
if the relation \eqref{eq41} is included into \eqref{eq42}. If we further define the electric field as $E(r) \coloneq c F_{0 1}$, it is straightforward to obtain from \eqref{eq41}, \eqref{eq43} and \eqref{eq33} that:

\begin{equation}
\label{eq44}
E(r) = \frac{1}{4 \pi \epsilon_{0}} \frac{Q_{L}}{r^{2}} \bigg(1 - \frac{r}{r_{L}}\bigg)^{2},
\end{equation}
which is a non monotonic function if $r_{L}$ is positive and with $Q_{L}/4\pi\epsilon_{0}r_{L}^{2}$ as its asymptotic limit for large radii, provided the Lyra scale radius is finite. 

The constant denoted as $Q_{L}$ in the electric field expression \eqref{eq44}, due to the form of the Maxwell-Lyra equations with source terms, can not be always trivially identified with the standard electric charge definition if the charge distribution is not punctual. Moreover, it is necessary to set $r_{0} = r_{L}$ in the Lyra scale function \eqref{eq41} to ensure that the same constant, differing only by a factor of $c$, appears in both \eqref{eq43} and \eqref{eq44}. In this manner, these two expressions directly converge to their well-established Lorentzian counterparts when $r_{L}$ approaches infinity and provided that $c_{1}$ is an appropriate constant.

Finally, substituting the Lyra scale function \eqref{eq41} and the electric potential \eqref{eq43} into the LyST equation \eqref{eq38} yields the $\alpha(r)$ expression, which results in the Lyra line element:
\begin{equation}
\label{eq45}
ds^{2} = \frac{r_{L}^{2}}{\Delta_{L}} 
\frac{\Delta(r)}{r^{2}} c^{2} dt^{2} - \phi^{4} \frac{\Delta_{L}}{r_{L}^{2}} \frac{r^{2}}{\Delta(r)}dr^{2} - \phi^{2} r^{2} d\Omega^{2}, 
\end{equation}
such that $\Delta(r) = (r - r_{+})(r - r_{-})$, $\Delta_{L} \coloneq \Delta(r_{L})$ and $\phi = (1 - r/r_{L})^{-1}$, with $r_{L}$, $r_{+}$ and $r_{-}$ being the roots of $\alpha(r)$. If we further set:
\begin{equation}
\label{eq46}
\begin{cases}
r_{s} = r_{+} + r_{-} \\
r_{Q}^{2} = r_{+} ~r_{-}
\end{cases},
\end{equation}
it is possible to define the Lyra geometrical mass $M$ and the Lyra geometrical charge $Q$ by considering:
\begin{equation}
\label{eq46.2}
\begin{cases}
2M \coloneq r_{s} r_{L}^{2} / \Delta_{L} \\
Q^{2} \coloneq r_{Q}^{2} r_{L}^{2} / \Delta_{L}
\end{cases}
\hspace{-0.3cm}
,
\end{equation}
so that, as a consequence, the metric tensor component $\tensor{g}{_0_0}$ can be rewritten as:
\begin{equation}
\label{eq47}
\alpha(r) = \phi(r)^{-2} \bigg[1 - \frac{2M}{r}\bigg(1 - \frac{r}{r_{L}}\bigg) + \frac{Q^{2}}{r^{2}}\bigg(1 - \frac{r^{2}}{r_{L}^{2}}\bigg)\bigg],
\end{equation} 
which can be shown to be consistent with the unused third LyST equation, a necessary step since we set \eqref{eq47} to describe the radial and temporal parts of the metric.

As for the relations in \eqref{eq46.2}, it is important to emphasize that the definition of charge provided above aligns with the one specified on expressions \eqref{eq43} and \eqref{eq44}, since the static electric potential was used to find \eqref{eq47} and the constant $r_{0}$ was fixed as $r_{L}$. Additionally, the geometrical mass $M$, which is equivalent to the Newtonian mass if $\phi = 1$, is consistent with the definition presented in reference \cite{Cuzinatto_2021}. In the absence of electric charges, i.e. $r_{Q} = 0$, the relationship between the Lyra mass and the roots of $\tensor{g}{_0_0}$ simplifies to that described in \cite{Cuzinatto_2021}.

\subsection{Properties} \label{subsec4.1}

The LyST theory is expected to reduce itself to the General Relativity form when the scale function is constant. Therefore, expression \eqref{eq47} becomes the Reissner-Nordström metric time component when $r_L \to \pm \infty$:
\begin{equation}
\label{eq48}
\lim_{r_L \to \pm \infty} \alpha(r) = 1 - \frac{2M}{r} + \frac{Q^{2}}{r^{2}},
\end{equation}
with the geometrical mass $M$ being now the standard Newtonian mass definition in natural units. Furthermore, the function \eqref{eq47} transforms into the spherical symmetric solution obtained by \cite{Cuzinatto_2021} if $Q = 0$, such that its root $r_{s} = (1/2M + 1/r_{L})^{-1}$ can be interpreted as the LyST generalization of the Schwarzschild radius \cite{Cuzinatto_2021}.

Another interesting limit is that of large distances from the source, the asymptotic expansion of the function \eqref{eq47} for $r \rightarrow \infty$ at leading order takes the form:
\begin{equation}
\label{eq49}
\alpha(r) = -\frac{\Lambda_{L}}{3} r^2,
\end{equation}
such that the constant in this expression is given by:
\begin{equation}
\label{eq50}
\Lambda_{L} = -\frac{3}{r_{L}^{2}}\bigg(1 + \frac{2M}{r_{L}} - \frac{Q^{2}}{r_{L}^{2}}\bigg).
\end{equation}
As a result, in analogy with the large distances limit of the Reissner-Nordström-de Sitter metric, expression \eqref{eq50} can be interpreted as an emerging cosmological constant. 

The mass and charge of objects defined on Lyra manifolds act as to produce a dark energy like effect if we solely consider the metric tensor components. However, due to the definition \eqref{eq13} in which $\phi^{2}$ appears multiplying $\tensor{g}{_\mu_\nu}$, the asymptotic $r \rightarrow \infty$ limit of the line element \eqref{eq45} will not behave as the de Sitter one. Moreover, it is possible to observe from expression \eqref{eq50} that the effective cosmological constant wil be positive (de Sitter type) if:
\begin{equation}
\label{eq51}
-M - \sqrt{M^{2} + Q^{2}} < r_{L} < -M + \sqrt{M^{2} + Q^{2}},
\end{equation}
and that \eqref{eq49} behaves as anti-de Sitter ($\Lambda_{L} < 0$) if: 
\begin{equation}
\label{eq52}
r_{L} > -M + \sqrt{M^{2} + Q^{2}} \ \text{or} \ r_{L} < - M - \sqrt{M^{2} + Q^{2}}.
\end{equation}

\subsection{Horizons and singularities} \label{subsec4.2}

With the metric of a charged black hole on the Lyra geometry defined in \eqref{eq45}, it is now essential to analyze the possible horizons and singularities of this solution. As it was revealed in the beginning of this section, there are three roots for $\alpha(r) = 0$. It can be easily found from \eqref{eq47} that a positive Lyra scale radius is one of them and that the other two can be written, by construction, with respect to $r_{s}$ and $r_{Q}$ as:
\begin{equation}
\label{eq53}
r_{\pm} = \frac{1}{2}\left(r_{s} \pm \sqrt{r_{s}^{2} - 4r_{Q}^2}\right).
\end{equation}

It is also important to understand for which cases $\alpha(r) \to \infty$. In this respect, there are two possible solutions:
\begin{equation}
\label{eq54}
\begin{cases}
r = 0  \\
{r_L}^2 - r_S r_L + {r_Q}^2 = 0
\end{cases},
\end{equation}
such that the latter can be viewed as a condition which requires $r_{\pm}$ and $r_{L}$ to be different. In addition, it is important to mention that for $r_{L} \rightarrow 0$ the field $\phi$ diverges for any finite value of $r$, therefore, this case is not considered in our analysis. To summarize, there are then four physical or coordinate singularities:
\begin{equation}
\label{eq55}
r = 0, ~ r = r_L \ \ \text{and} \ \ r = r_\pm.
\end{equation}

\subsubsection{Physical singularities} \label{subsubsec4.2.1}

To ensure that some of the values of $r$ defined in \eqref{eq55} are physical singularities it is essential to understand the behavior of curvature scalar invariants on these points. The Ricci scalar of Lyra geometry, relation \eqref{eq26}, vanishes at every point since the solution in consideration is obtained in the absence of matter fields. For this reason, it is necessary to consider the Kretschmann scalar $K = R_{\alpha \beta \mu \nu } R^{\alpha \beta \mu \nu }$, with the Riemann tensor given by \eqref{eq22}, so that the expression:

\begin{eqnarray}
\label{eq56}
\hspace{-0.7cm}
K &=& \frac{8}{r^{8}} \left(1 - \frac{r}{r_{L}}\right)^{6}  \\
&& \times \bigg[Q^{4} \left(7 - \frac{2r}{r_{L}} + \frac{r^{2}}{r_{L}^{2}}\right) + 6 M r \left(M r - 2 Q^{2}\right) \bigg], \nonumber
\end{eqnarray}
in which \eqref{eq46.2} was utilized, reveals that at $r = 0$ a physical singularity certainly exists, since this curvature scalar diverges at this radius. Additionally, as expected, the General Relativity expression is recovered for $r_L \to \pm \infty$:
\begin{equation}
\label{eq57}
\lim_{r_L \to \pm \infty} K = \frac{56 Q^{4}}{r^{8}} - \frac{96 M Q^{2}}{r^{7}} + \frac{48 M^{2}}{r^{6}}.
\end{equation}

Furthermore, as the Ricci scalar is zero everywhere, the Kretschmann scalar becomes equivalent to the contraction of the Weyl tensor with itself, commonly known as the Weyl curvature scalar. Due to the relation between this tensor and the shear of a geodesic congruence, $\sqrt{K}$ can be interpreted as a measure of the intensity of tidal deformations at a given radius. As an example, $K$ vanishes at $r = r_{L}$, so that this radius can be characterized by a null tidal force intensity and a Lyra field $\phi$ that diverges. Nevertheless, a geodesic analysis is further necessary to comprehend more about $r = r_{L}$, as is also the case for $r = r_{\pm}$.

\subsubsection{Horizons} \label{subsubsec4.2.2}

In the Reissner-Nordström spacetime obtained from General Relativity it is well-known that two important null hypersurfaces exist: the inner and exterior horizons, also respectively referred as the Cauchy and event horizons. These two surfaces are defined from relations \eqref{eq53}, so that in this case $r_{s}$ is only related to the mass and $r_{Q}$ to the charge, since for $\phi = 1$ the quantity $\Delta_{L}$ is equal to $r_{L}^{2}$ in \eqref{eq46.2}. As a consequence, since this spacetime solution is a particular case of the metric obtained from the function \eqref{eq47}, it is \textit{a priori} reasonable to designate $r_{+}$ as the exterior horizon and $r_{-}$ as the inner one.

Nonetheless, the relation between these horizons and the Lyra geometrical mass and charge is not so trivial as in the case of a constant $\phi$. The presence of an effective cosmological constant alters this relation when compared to the aforementioned particular case. As a result, by considering the relations in \eqref{eq46.2}, the expressions \eqref{eq53} take the form:
\begin{equation}
\label{eq58}
r_{\pm} = \frac{M \pm \sqrt{M^{2} - \lambda_{L} Q^{2} }}{\lambda_{L}},
\end{equation}
such that this new factor $\lambda_{L}$, which depends on the Lyra radius and on the mass and charge of the black hole, is related to \eqref{eq50} via:
\begin{equation}
\label{eq59}
\lambda_{L} = - \frac{\Lambda_{L}}{3} r_{L}^{2}.
\end{equation}

The quantity inside the square root of relation \eqref{eq58}, which from now on is designated as $\Xi_{L} \coloneq M^{2} - \lambda_{L} Q^{2}$, and the factor $\lambda_{L}$ defined in \eqref{eq59}, are fundamental to the comprehension of the LyST charged black hole anatomy. These two expressions are essential to understand the inner and exterior horizons behaviors when considering different mass and charge configurations. For this purpose, it is necessary to consider the values of $Q$ for which $\Xi_{L} = 0$:
\begin{equation}
\label{eq60}
\begin{cases}
_{+}Q^{\text{x}}_{\pm} = +\frac{1}{2}\left(r_L \pm \sqrt{{r_L}^2 + 4 M r_{L}}\right) \\
_{-}Q^{\text{x}}_{\pm} = - _{+}Q^{\text{x}}_{\pm}
\end{cases}
\hspace{-0.3cm}
,
\end{equation}
and the roots of $\lambda_{L}$, which are expressed as:
\begin{equation}
\label{eq61}
Q_\pm = \pm \sqrt{r_{L}^2 + 2 M r_{L}}.
\end{equation}

\begin{figure*}[htb]
\centering
\includegraphics[scale=0.47]{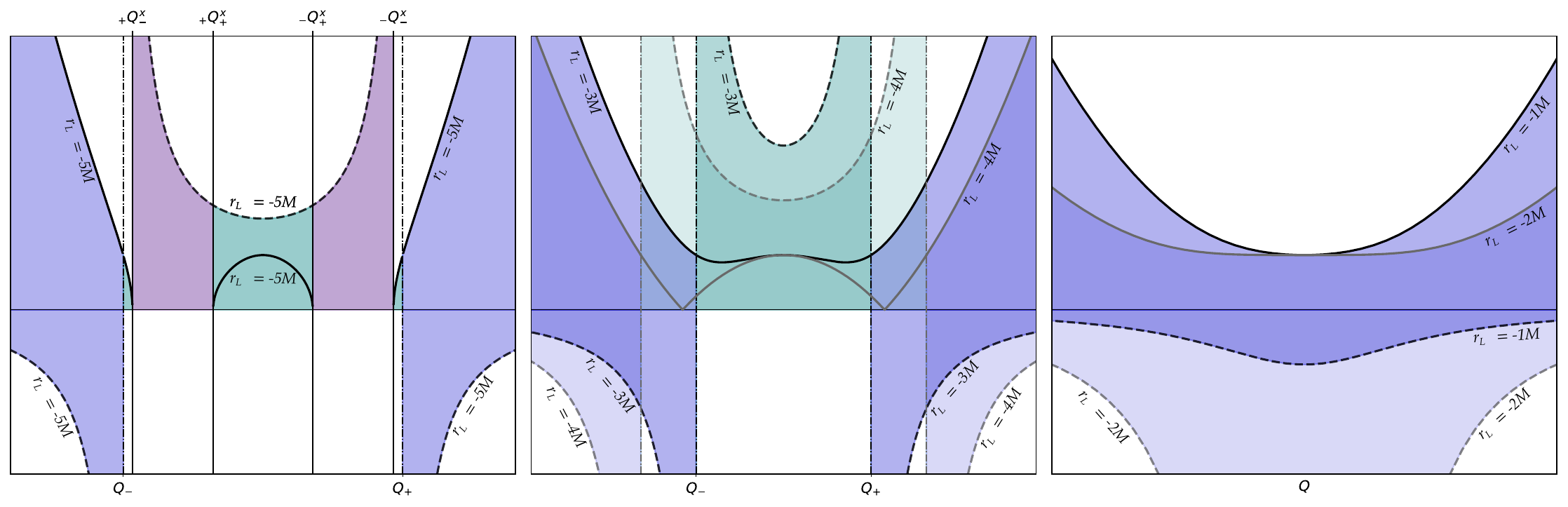}
\caption{{\it Left:} $r_L < -4M$. {\it Center:} $-4M \leq r_L < -2M$. {\it Right:} $-2M \leq r_L < 0$. The black solid curves represent $\Xi_{L}^{1/2}$ as function of $Q$ for different values of $r_{L}$ and the dashed ones correspond to $1/\lambda_{L}$. The black vertical continuous lines are associated with the extremal charges and the dashdotted ones with the roots $Q_{\pm}$. It is also shown the limit cases (gray curves) $r_{L} = -4M$ and $r_{L} = -2M$. The colored regions just represent the charge intervals in which they are confined: \textit{Green:} two horizons, \textit{Blue:} one horizon, \textit{Indigo:} naked singularity.}
\label{fig:horizons}
\end{figure*}

\begin{figure}[htb]
\centering
\hspace{-0.22cm}
\includegraphics[scale=0.395]{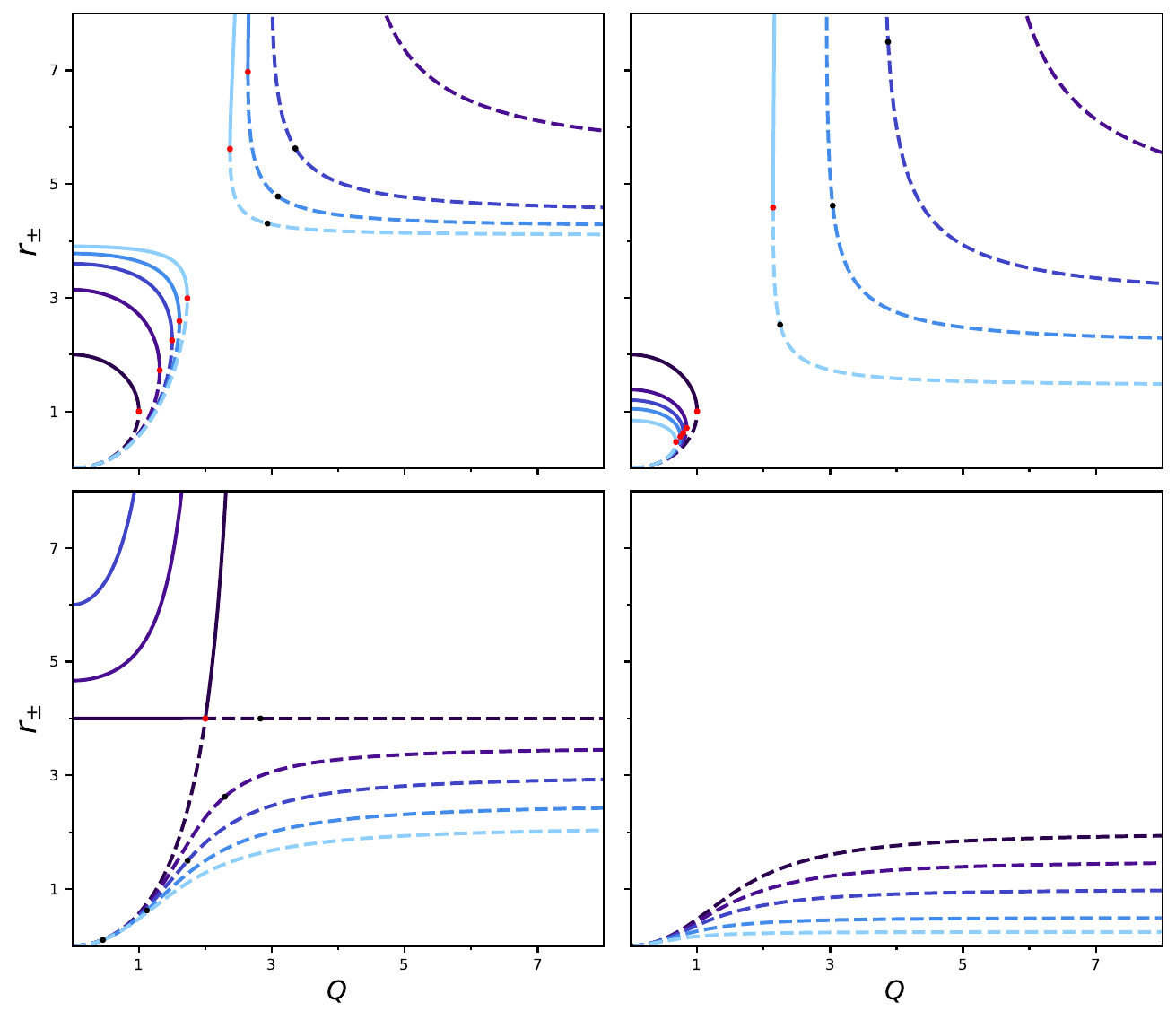}
\caption{{\it Top Left:} $r_L < -4M$. {\it Top Right:} $r_L > 0$. {\it Bottom Left:} $-4M \leq r_L < -2M$. {\it Bottom Right:} $-2M \leq r_L < 0$. The continuous lines represent $r_{+}$ and the dashed ones $r_{-}$, each line corresponds to a value of $r_{L}$. The blue curves become darker as $\abs{r_{L}}$ increases, for $M = 1$, the scale radius values utilized were: $(-4.1, -4.25, -4.5, -5.5, -\infty)$, $(1.46, 2.2, 3, 4.5, \infty)$, $(-2.1, -2.5, -3, -3.5, -4)$ and $(-0.25, -0.5, -1, -1.5, -2)$. The red dots describe the extremal charges $Q = {}_{\pm}Q^{\text{x}}_{\pm}$, the black ones represent $Q = Q_{\pm}$. The darker colors were utilized for the $r_L \to \pm \infty$, $r_L = -4M$ and $r_L = -2M$ cases.}
\label{fig: event and cauchy horizons}
\end{figure}

The relations in \eqref{eq60} have information as to which charge values $r_{+}$ equates to $r_{-}$ and as to when these roots become complex numbers. As for the mathematical expressions in \eqref{eq61}, they describe for which values of $Q$ the radius $r_{+}$ is non-negative. As a consequence, when examining the conditions under which the square roots of these expressions vanish, it becomes natural to define three distinct regimes: $r_L > 0$ or $r_L < -4 M$, $-4M \leq r_L < -2M$ and $-2M \leq r_L < 0$. Depending on the charge value $Q$ and on the regime within which $r_{L}$ is located, there will be four different types of physical objects: one solution defined by two horizons, two peculiar varieties of extremal black holes and a naked singularity. These possibilities are shown in Figure \ref{fig:horizons} and \ref{fig: event and cauchy horizons}.

If the horizons configurations are thoroughly scrutinized, it is interesting to observe that for the first regime, for which:
\begin{itemize}
\item $r_{L} > 0$ or $r_{L} < -4M$, 
\end{itemize}
the horizons will cease to exist, i.e. $r_{\pm} \in \mathbb{C}$, if for $r_{L} > 0$:
\begin{equation}
\label{eq62.1}
Q \in ({}_{-}Q^{\text{x}}_{+}, {}_{+}Q^{\text{x}}_{-}) \ \ \text{or} \ \ Q \in ({}_{-}Q^{\text{x}}_{-}, {}_{+}Q^{\text{x}}_{+}), 
\end{equation}
and for $r_{L} < - 4M$ if the charge value satisfies:
\begin{equation}
\label{eq62.2}
Q \in ({}_{+}Q^{\text{x}}_{-}, {}_{+}Q^{\text{x}}_{+}) \ \ \text{or} \ \ Q \in ({}_{-}Q^{\text{x}}_{+}, {}_{-}Q^{\text{x}}_{-}), 
\end{equation}
resulting in the emergence of a naked singularity on those cases. As a consequence, an extremal black hole with $r_{+} = r_{-}$ will form if $Q = {}_{\pm}Q^{\text{x}}_{\pm}$, such that its radius is defined by:
\begin{equation}
\label{eq62}    
\lim_{Q \to {}_{\pm}Q^{\text{x}}_{\pm}} r_{\pm} = \frac{2 M}{1 + \frac{2 M}{r_{L}} \mp \frac{\sqrt{r_{L}^2 + 4 M r_{L}}}{r_{L}}}, 
\end{equation}
in which the upper sign of $\mp$ corresponds to $Q = {}_{\pm}Q^{\text{x}}_{+}$. In addition, the exterior horizon $r_{+}$ will have a negative value if $Q > Q_{+}$ or $Q < Q_{-}$, so that an extremal black hole with $r_{-}$ as its horizon emerges. Therefore, in this regime, if for $r_{L} > 0$ the charge is in the open interval:
\begin{equation}
\label{eq62.3}
\hspace*{-0.2cm}
(Q_{-}, {}_{-}Q^{\text{x}}_{+}), \ (_{+}Q^{\text{x}}_{-}, {}_{-}Q^{\text{x}}_{-}) \ \ \text{or} \ \ ({}_{+}Q^{\text{x}}_{+}, Q_{+}),
\end{equation}
and if for $r_{L} < -4M$ it can be found in:
\begin{equation}
\label{eq62.4}
\hspace*{-0.2cm}
(Q_{-}, {}_{+}Q^{\text{x}}_{-}), \ (_{+}Q^{\text{x}}_{+}, {}_{-}Q^{\text{x}}_{+}) \ \ \text{or} \ \ ({}_{-}Q^{\text{x}}_{-}, Q_{+}),
\end{equation}
the LyST charged black hole will possess two horizons. 

This transition between the different types of horizons configurations can be better perceived if the first graphic of Figure \ref{fig:horizons} is considered. It can be observed that as the charge $Q$ increases from negative to positive values, the first and last possibilities are characterized by a black hole with one horizon, i.e. $r_{-}$, which is represented by the blue colored regions encompassing the values of $Q$ for which $r_{+} < 0$. Between these cases, there is an alternation among green regions, for which there are two horizons, and the indigo ones, which symbolize the naked singularity cases.

Moreover, the top graphics of Figure \ref{fig: event and cauchy horizons} also show these considerations. It can be observed, from $Q = 0$ to $Q = {}_{-}Q^{\text{x}}_{-} (\text{or} {}_{-}Q^{\text{x}}_{+})$, that the horizons behave in a similar manner as the Reissner-Nordström ones, such that the $r_{+} = r_{-}$ black hole for the General Relativity case forms at:
\begin{equation}
\label{eq63}
\lim_{r_L \to \pm \infty} {}_{-}Q^{\text{x}}_{-} = M.
\end{equation}
Thus, considering \eqref{eq60}, the equality between the mass and charge on a Lyra manifold with finite scale radius does not generate an extremal black hole. As a consequence, in contrast to the infinite $r_{L}$ case, black holes with $Q > M$ are conceivable in this physical scenario. Besides, it can be further seen from these graphics, that the possibility of a resurgence of the horizons after the naked singularity emergence is non-null. 

If this phenomenon is physically plausible, which will be considered in section \ref{sec5}, then, as the charge increases, a horizon with $r_{+} = r_{-}$ first arises covering the naked singularity, it eventually splits into $r_{+}$ and $r_{-}$, so that, for $Q$ growing beyond $Q_{+}$, the exterior horizon $r_{+}$ simply vanishes. On those cases in which $r_{\pm}$ are real numbers, the interior horizon never disappears, since its numerator has always the same sign as its denominator, resulting in $r_{-} > 0$. Furthermore, in this first regime, at values of $Q$ equal to the charge values in \eqref{eq61}, the $r_{+}$ horizon diverges and becomes discontinuous and the inner one reaches the limit:
\begin{equation}
\label{eq64}
\lim_{Q \to Q_{\pm}} r_{-} = r_L \left(1 + \frac{r_{L}}{2M} \right),
\end{equation}
so that as $Q$ goes to infinity, the horizons attain the limits:
\begin{equation}
\label{eq65}
\lim_{Q \to \pm \infty} r_{\pm} \to \mp \abs{r_{L}}.
\end{equation}

As for the mentioned second interval of the Lyra scale radius, that is:
\begin{itemize}
\item $-4M \leq r_L < -2M$, 
\end{itemize}
the extremal charges in \eqref{eq60} reduce to $\pm r_{L}/2$ if $r_{L} = -4 M$, so that the possibility of a naked singularity formation is null. The extremal black hole is then defined by \eqref{eq62} with its square root term vanishing, which results in a extremal horizon given by the Schwarzschild-Lyra radius mentioned in subsection \ref{subsec4.1}. In this particular case, therefore, there will be two horizons if $Q \in (Q_{-}, Q_{+})$ and $Q \neq \pm r_{L}/2$. For the rest of the interval, the extremal charges \eqref{eq60} become complex numbers, so that only the extremal $r_{+} < 0$ black hole will form, if $Q > Q_{+}$ or $Q < Q_{-}$. As a consequence, this regime has at least one everlasting horizon.

For the third and last scale radius and mass configuration, defined by:
\begin{itemize}
\item $-2M \leq r_L < 0$, 
\end{itemize}
there is no extremal charges since ${}_{\pm}Q^{\text{x}}_{\pm} \in \mathbb{C}$. It can be further noted that $Q_{\pm} \in \mathbb{C}$ or, if $r_{L} = -2 M$, that $Q_{\pm} = 0$. As a result, the exterior horizon $r_{+}$ has always negative values in this regime and there is no charge intervals in which the appearence of a naked singularity is possible. If the case in which $r_{+} = 0$, i.e. $r_{L} \rightarrow - 2 M$, is to be regarded as unphysical, this regime is then characterized by the solely presence of the $r_{-}$ horizon. This behavior can be better understood in the bottom right graphic of Figure \ref{fig: event and cauchy horizons}, in which this horizon reaches the limit $\abs{r_{L}}$ as $Q \rightarrow \infty$.

\subsection{Geodesics and causal structure} \label{subsec4.3}

To properly determine if the surfaces $r_{\pm}$ defined by \eqref{eq58} are apparent horizons, it is essential to understand the behavior of null and time-like geodesics. By fixing the angular coordinates $\theta = \pi/2$ and $\dot{\varphi} = 0$, the only necessary remaining Lyra geodesic equation from \eqref{eq14} can be expressed in terms of $r_{\pm}$ as:
\begin{equation}
\label{eq66}    
\frac{dt}{d\tau} = \frac{k}{\alpha \phi^{2}} = k \frac{\Delta_{L}}{r_{L}^{2}} \frac{r^{2}}{\Delta(r)},
\end{equation}
in which $\tau$ is an affine parameter and $k$ an \textit{a priori} positive real constant. This relation tends to positive infinity if $r \rightarrow r_{+}^{+}$, so that, as is typical of apparent horizons, any information transmitted by an infalling observer from $ r_{-} < r \leq r_{+}$ to $r > r_{+}$ will be infinitely redshifted. However, this behavior is inverted if the peculiar $r_{-} < r_{L} < r_{+}$ case is considered.

Similar to what happens at the Cauchy horizon of the Reissner-Nordström metric, since \eqref{eq66} tends to $- \infty$ if $r \rightarrow r_{-}^{+}$, an infalling observer at $r = r_{-}$ will receive infinitely blueshifted radiation from $r > r_{-}$ if the scale radius does not satisfy $r_{-} < r_{L} < r_{+}$. Nevertheless, since it was revealed at subsection \ref{subsec4.2} that the exterior horizon fade away for certain charge values and that, as will be shown, the metric necessarily changes its causal nature in $r < r_{-}$, the inner horizon will act as an usual apparent horizon if $r_{+} < 0$, given that \eqref{eq66} generally tends to $+\infty$ for $r \rightarrow r_{-}^{+}$ in this particular scenario. However, it will oddly tend to negative infinity if $0 < r_{L} < r_{-}$ or, for negative values of $r_{L}$, if $r_{L} > r_{+}$.

In order to analyze how non-massive particles are radially affected by the LyST charged black hole, it is assumed the normalization $u_{\mu} u^{\mu} = 0$, with the four-velocity defined by \eqref{eq5}. As a result, the equation:
\begin{equation}
\label{eq67}    
\frac{dt}{dr} = \pm \frac{1}{\alpha} = \pm \left(1 - \frac{r}{r_{L}}\right)^{-2} \frac{\Delta_{L}}{r_{L}^{2}} \frac{r^{2}}{\Delta(r)},
\end{equation}
is obtained, in natural units, through the use of \eqref{eq66}. Therefore, for a constant $t_{0}$, the coordinate time measured for null geodesics as a function of $r$ can be expressed as:
\begin{eqnarray}
\label{eq68}
\hspace*{-0.6cm}
t_\pm &=& t_{0} \pm \Bigg\{r \phi + r_L \left( \frac{r_{+}}{r_{L} - r_{+}} + \frac{r_{-}}{r_{L} - r_{-}}\right) \ln{\phi} \\
&& + \frac{\Delta_{L}}{r_{+} - r_{-}} \left[\frac{r_{+}^{2} \ln{(r - r_{+})}}{(r_{L} - r_{+})^{2}} - \frac{r_{-}^{2} \ln{(r-r_{-})}}{(r_{L} - r_{-})^{2}}\right] \Bigg\}, \nonumber
\end{eqnarray}
with the scale field given by \eqref{eq41} for $r_{0} = r_{L}$. As expected, relation \eqref{eq68} results in the General Relativity case for an infinite Lyra scale radius.

\begin{figure}[htb]
\centering
\includegraphics[scale=0.388]{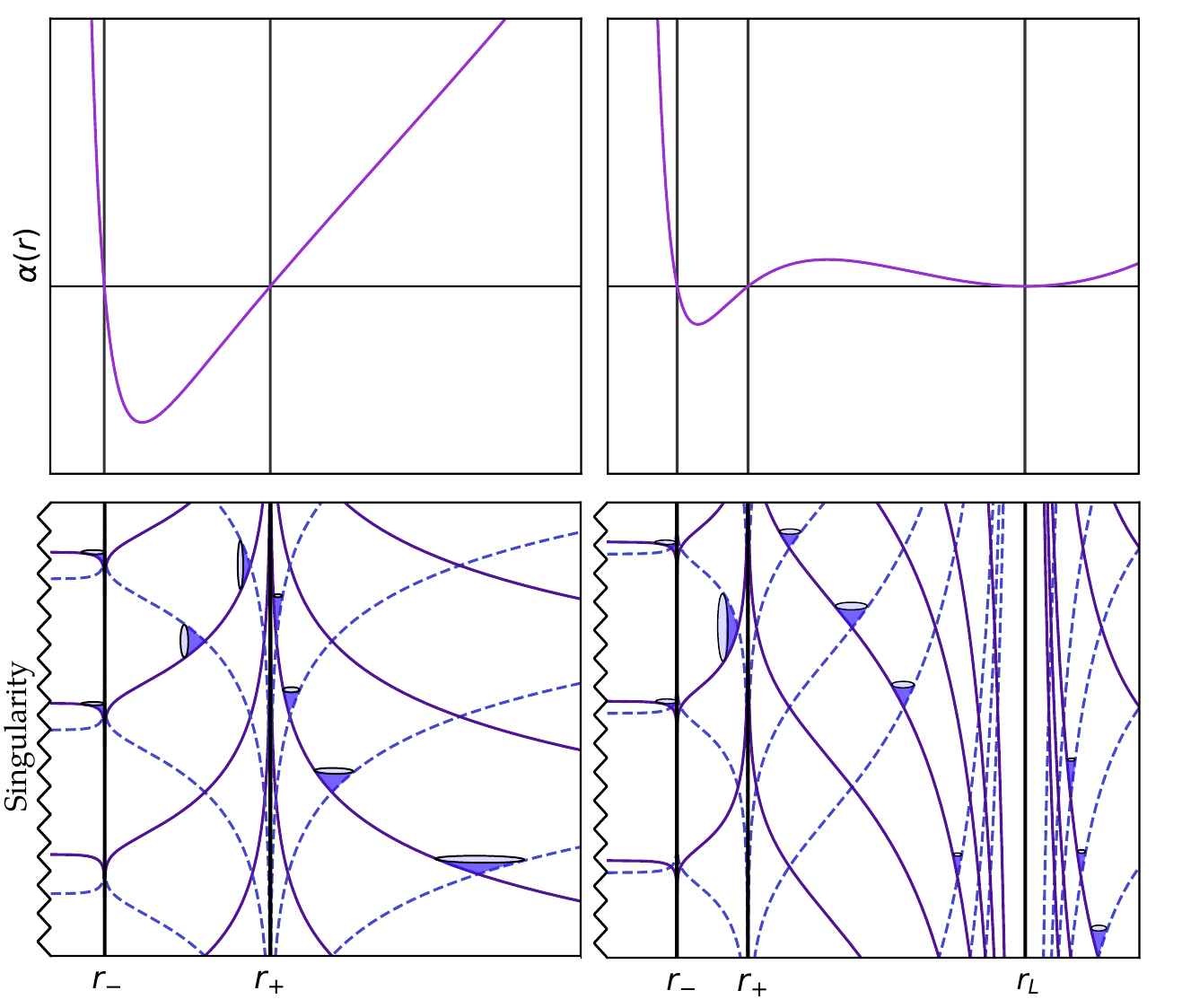}
\caption{{\it Left side:} $r_{L} < 0$. {\it Right side:} $r_{L} > 0$. In the top graphics, the plotted curves correspond to $\alpha(r)$. In the bottom ones, the coordinate time solutions $t_{+}$ (light blue dashed lines) and $t_{-}$ (dark blue solid lines) of null geodesics are shown as functions of $r$. Future directed null cones are artistically represented. The vertical black solid lines represent $r = r_{-}$, $r = r_{+}$ and $r = r_{L}$.}
\label{fig: geodesics}
\end{figure}

The behavior of the metric component \eqref{eq47} and of ingoing and outgoing null geodesic congruences defined by \eqref{eq68} are shown in Figure \ref{fig: geodesics}. In the top graphics, it is straightforward to notice that $\alpha(r)$ is negative for $r \in (r_{-}, r_{+})$, so that the coordinate time $t$ becomes spacelike and $r$ assumes a timelike character within this non-static region. Consequently, in the remaining radial intervals, these coordinates return to their conventional causal configurations. Thus, null cones are rotated by a right angle only between both horizons or, for the cases in which the exterior horizon is absent, if $r < r_{-}$. It is also possible to observe, in the bottom graphics, that null geodesics reach $r = r_{+}$ at an infinite coordinate time, so that light cones are continuously squeezed for $r \rightarrow r_{+}^{+}$. As a result, the causal structure of this spacetime is qualitatively similar to the Reissner-Nordström one.

For the $S^{2}$ sphere defined by a positive Lyra radius, however, the coordinate $t$ is timelike inside and outside it. Moreover, expression \eqref{eq66} indicates no infinite redshift possibility in this region. Thus, although $t_{\pm}$ diverges at $r = r_{L}$ from both directions, the region $r = r_{L}$ does not seem to be an apparent horizon. It is, thereby, necessary to consider the trajectory of free falling massive particles at this radius. For this purpose, the normalization $u_{\mu} u^{\mu} = 1$, in natural units, leads to the equation:
\begin{equation}
\label{eq69}
\frac{d\tau}{dr} = \pm \bigg(1 - \frac{r}{r_{L}}\bigg)^{-2} \frac{\sqrt{\Delta_{L}}}{\abs{r_{L}}}\bigg(\frac{\Delta(r_{0})}{{r_{0}}^{2}} - \frac{\Delta(r)}{r^{2}}\bigg)^{-1/2},
\end{equation}
such that for $r = r_{0}$ the right side of the equation above vanishes.

\ 

\begin{figure}[htb]
\centering
\includegraphics[width=4cm]{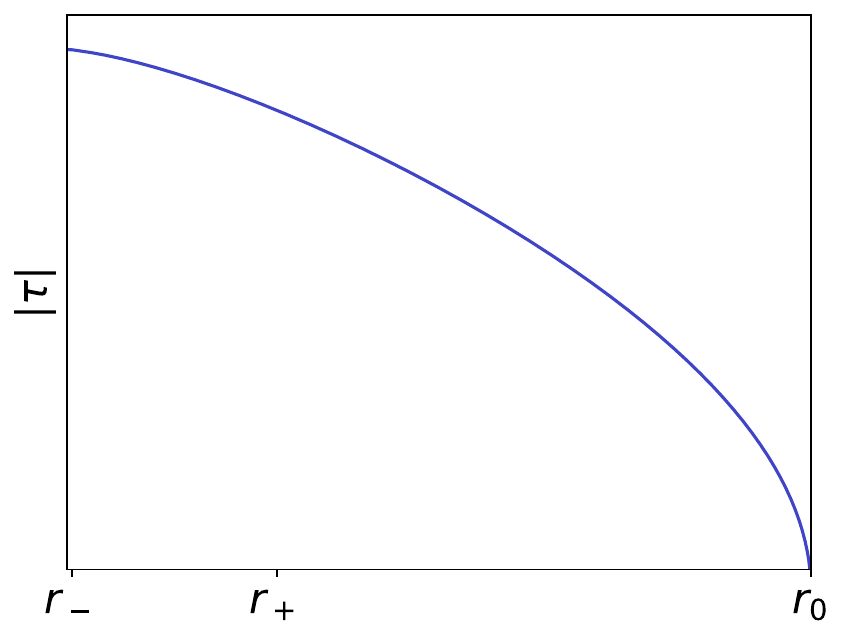}%
\includegraphics[width=4cm]{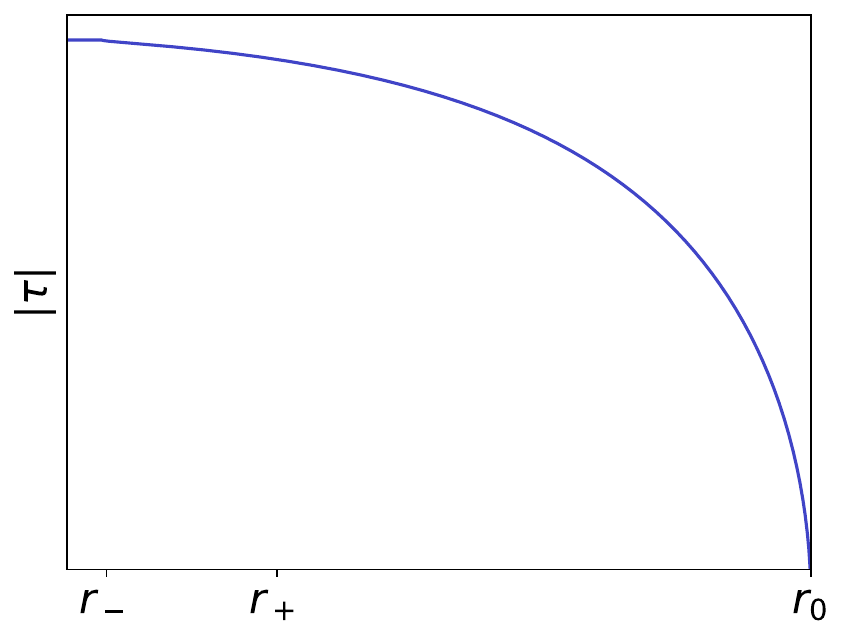}
\includegraphics[width=4cm]{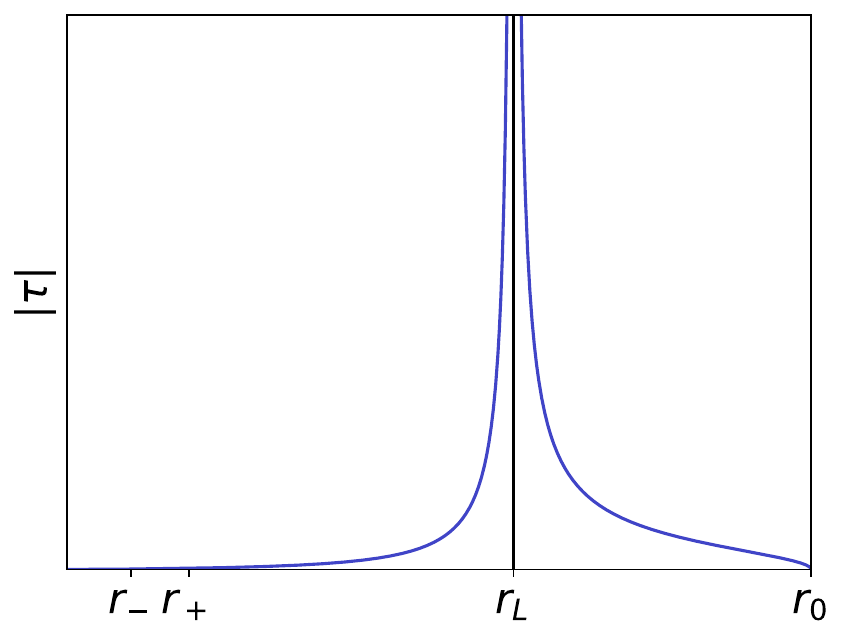}
\caption{{\it Left:} $r_{L} < 0$. {\it Right:} $r_{0} < r_{L}$. {\it Bottom:} $0 < r_{L} < r_0$. The blue curves represent the absolute value of massive particles proper time plotted as function of $r$. Different configurations of the initial position $r_{0}$ are considered. The black vertical solid line represents the positive $r_{L}$ case.}
\label{fig: proper time geodesics}
\end{figure} 

\

Despite the fact that expression \eqref{eq69} possesses an analytic solution, its mathematical expression is excessively large to be present in this current text. Nonetheless, its behavior in different configurations is shown in Figure \ref{fig: proper time geodesics}. From the bottom graphic in this figure and by examining expression \eqref{eq69}, it becomes evident that the proper time of a massive particle tends to infinity as it approaches the Lyra surface defined by $r_{L} > 0$. A particle coming from $r_0 > r_{L}$ takes an infinite proper time to reach $r = r_{L}$ or to move out of it and fall in the black hole. As a consequence, a positive Lyra scale radius separates the spacetime manifold into two regions that no non-accelerated observers can trespass. 

For observers trying to travel through the surface defined by $r_{L}$ it would be as if they were venturing into spatial infinity, an examination of the black hole physics would then be essential to characterize the corresponding geometry. The fact that tidal deformations are null at the Lyra radius, the proper time $\tau$ of free falling observers increases infinitely as $r$ approaches $r_{L}$ and the coordinate time $t$ is equivalent to the proper time of a fixed observer at the surface $r_{L}$ (see \eqref{eq45}, for example), is a further indication that the $r = r_{L}$ case is equivalent to the scenario of infinitely distant regions of conventional Riemannian manifolds. In other words, one of the effects of the Lyra field $\phi$ is to compactify spacetime. As a result, for a non-negative scale radius, the radial coordinate $r$ must satisfy $r \in [0, r_{L}]$ and $r_{L} > r_{\pm}$ for a plausible physical solution. 

\section{Overcharging}
\label{sec5}

In the previous section, it was shown that Lyra charged black holes can exist for $|Q| > M$ and that a timelike naked singularity can arise for certain charge intervals. As for the current section, an investigation is employed to assess the possibility of overcharging LyST black holes defined by the metric \eqref{eq45} as to expose the physical singularity. In special, an attempt of overcharge is also made to skip the charge intervals in which the horizons vanish, so that an eternal charged black hole can emerge. In both scenarios, it is considered the phenomenological case in which $r_{L} > 0$ or $r_{L} < -4M$.

The following analysis is a generalization of the one constructed for Reissner–Nordström black holes by reference \cite{Hubeny1999}. As is done by the mentioned work, it is considered $Q > 0$ without loss of generality. The first step to expose the singularity is to consider a near-extremal charged black hole. A particle with charge $q$, rest mass $m$ and energy $E$ is radially sent to the black hole satisfying $m < E < q \ll Q < {}_{-}Q^{\text{x}}_{\pm}$, in which ${}_{-}Q^{\text{x}}_{-}$ corresponds to $r_{L} > 0$ and ${}_{-}Q^{\text{x}}_{+}$ to $r_{L} < -4M$ (see relations \eqref{eq62.1} and \eqref{eq62.2}). Three conditions are then necessary: the particle descends past the exterior horizon, the final configuration overcomes the first extremal scenario without surpassing the second one, that is $Q + q >{}_{-}Q^{\text{x}}_{\pm} + E$ such that $Q + q <{}_{+}Q^{\text{x}}_{+} (\text{or} \ {}_{-}Q^{\text{x}}_{-} \ \text{if} \ r_{L} < -4M) + E$, and a test particle is to be considered as the overcharge causing agent, so that backreaction effects are negligible.

The restriction on the maximum allowable charge for the black hole's final state, wherein $Q + q$ must not reach or exceed ${}_{+}Q^{\text{x}}_{+} + E$ or ${}_{-}Q^{\text{x}}_{-} + E$, is naturally satisfied when $r_{L} > 0$ or $r_{L} < -4M$, given that a test charged particle is considered. In essence, this requirement is satisfied due to the particle's energy condition $E < q \ll Q$ and the fact that for values of $r_{L}$ not too close to zero or $-4M$ the charge interval between the first and second extremal charges is large enough (see the first plot in Figure \ref{fig:horizons}). If otherwise, backreaction effects would be non-negligible, and the mentioned requirement might not be satisfied. 

As for the second analysis, the black hole final charge is required to surpass the charge interval in which the horizons vanish. Therefore, the only difference to the analysis of the first case is the presence of the black hole final state requirement: $Q + q >{}_{+}Q^{\text{x}}_{+} + E$ when $r_{L} > 0$ or $Q + q >{}_{-}Q^{\text{x}}_{-} + E$ for $r_{L} < -4M$. However, to maintain the approach of charged test particles, the charge interval in which a naked singularity is possible has to be small enough so that a slight overcharge has the potential to yield an eternal black hole. As a result, it is further required that the Lyra radius value is close to $-4M$ or $0$.

To perform both analyses, it is convenient to define a new coordinate $v = t + r_{*}$, such that $dr_{*} = \alpha^{-1} dr$. The black hole metric \eqref{eq45} is then rewritten as:
\begin{equation}
\label{eq70}
ds^{2} = \phi^{2} \left( \alpha dv^{2} - 2 dv dr - r^{2} d\Omega^{2} \right),
\end{equation}
for $\alpha$ given by the solution \eqref{eq47}. Furthermore, it is important to note that since a scale transformation is not performed and the coordinate one is just given by $x^{\mu} = (t,r,\theta,\varphi) \rightarrow \bar{x}^{\mu} = (v,r,\theta,\varphi)$, the transformation law \eqref{eq9} applied to the electric potential yields $\bar{A}_{\mu} = A_{\mu}$.

To describe the test charged particle dynamics it is utilized the equation of motion:
\begin{equation}
\label{eq71}    
\bar{u}^{\alpha} \nabla_{\alpha} \bar{u}^{\mu} = \frac{q}{m} \tensor{\bar{F}}{^\mu_\beta} \bar{u}^{\beta},
\end{equation}
in which relation \eqref{eq33} is considered and such that the four-velocity components are defined by \eqref{eq5}. The lagrangian that generates the above equation is constructed as $L = \tensor{\bar{g}}{_\mu_\nu} \bar{u}^{\mu} \bar{u}^{\nu}/2 + q \bar{A}_{\mu} \bar{u}^{\mu}/m$, which can be rewritten for radial trajectories, $\dot{\theta} = \dot{\varphi} = 0$, as:
\begin{equation}
\label{eq72}
L = \frac{1}{2} \phi^2 \left[ \alpha \dot{v}^2 - 2 \dot{v} \dot{r} + \frac{2qQ}{m r} \phi^{-2} \dot{v} \right],
\end{equation}
in which the dot symbol represents the derivative with respect to the particle's proper time and such that $c_{1} = 0$ in \eqref{eq43}. Thus, since the Euler-Lagrange equations are valid in this case and \eqref{eq72} does not explicitly depend on $v$, the canonical momentum conjugate to this new coordinate is the particle's energy:
\begin{equation}
\label{eq73}
\frac{E}{m} \equiv \frac{\partial L}{d \dot{v}} =  \phi^{2}\left(\alpha \dot{v} - \dot{r} \right) + \frac{q Q}{m r}.
\end{equation}

It is also considered a timelike trajectory for the charged particle, such that $\bar{u}^{\mu} \bar{u}_{\mu} = 1$, which leads to:
\begin{equation}
\label{eq74}
\alpha \dot{v}^{2} - 2 \dot{v} \dot{r} = \phi^{-2}.
\end{equation}
Isolating $\dot{v}$ in the particle's energy relation \eqref{eq73} and substituting it into the above equation yields: 
\begin{equation}
\label{eq75}
\dot{r}^2 = \frac{1}{m^2 \phi^4} \left( E - \frac{q Q}{r} \right)^2 - \frac{\alpha}{\phi^2}, 
\end{equation}
which must obey the condition $\dot{r}^{2} > 0$ when $r \geq r_{+}$, since $\dot{r} = 0$ is a turning point, as to guarantee that the charged particle falls into the black hole. Using \eqref{eq75}, this condition leads to: 
\begin{equation}
\label{eq76}
\left(E - \frac{q Q}{r} \right)^{2} > m^{2} \phi^{2} \alpha.
\end{equation}
Given that $\phi^{2} \alpha(r)$ is non-negative for values of $r$ greater than $r_{+}$ when $r_{L} > r_{+}$ or $r_{L} < - 4M$, the inequality in \eqref{eq76} is equivalent to $E > qQ/r$. As a result, since $\dot{r}^{2} > 0$ for $r \geq r_{+}$ \cite{Hubeny1999}, the lower bound on the particle's energy is simply given by:
\begin{equation}
\label{eq77}
E > \frac{q Q}{r_{+}}.
\end{equation}

\subsection{Exposing the singularity} \label{subsubsec5.1}

To obtain a naked singularity it is necessary to consider the second mentioned condition, in which the final black hole charge must surpass the first extremal case, that is, $Q_{\text{final}} >{}_{-}Q^{\text{x}}_{\pm} + E$. Therefore, the upper bound for the energy is:
\begin{equation}
\label{eq78}
E < Q -{}_{-}Q^{\text{x}}_{\pm} + q.
\end{equation}
These two energy bounds imply that a LyST extremal black hole can be overcharged, since the extremality condition doesn't imply in $Q = r_{+} = M$, as is the case of the Reissner-Nodström metric, in which is not possible to overcharge an extremal solution \cite{Hubeny1999}. Moreover,
if the expressions \eqref{eq77} and \eqref{eq78} are combined, an inequality for the particle's charge is obtained:
\begin{eqnarray}
\label{eq79}
q &>& \left(1 - \frac{Q}{r_+} \right)^{-1} \left({}_{-}Q^{\text{x}}_{\pm} - Q \right).
\end{eqnarray}

It is further necessary to address the particle's mass values that satisfy $\dot{r}^{2} > 0$. By considering \eqref{eq76}, it is easy to find that $m < \phi^{-1} \alpha^{-1/2} \left(E - q Q r^{-1}\right)$ for $r \geq r_{+}$. Calculating the minimum value of the function in the right-hand side of this relation for $r = r_{m}$:
\begin{equation}
\label{eq80}
r_{m} = \frac{1}{2} r_{s} + \frac{1}{6} \left(3^{1/3} \mathcal{A}^{1/3} - 3^{2/3} \bar{\alpha} \mathcal{A}^{-1/3} \right),
\end{equation}
in which $\mathcal{A} = \bar{\beta} + \sqrt{3\bar{\alpha}^3 + \bar{\beta}^2}$ with $\bar{\alpha}$ and $\bar{\beta}$ given by:
\begin{equation}
\label{eq81}
\begin{cases}
\bar{\alpha} = 2 \bar{\gamma} - 3 {r_{s}}^2, \\
\bar{\beta} = 36 r_{L} {r_{Q}}^{2} - \bar{\gamma} r_{s} + {r_{s}}^3, \\
\bar{\gamma} = {r_Q}^2 + r_{L} r_{s} - \frac{q Q}{E} \left(2 r_{L} - r_{s} \right).
\end{cases}
\end{equation}

As a result, utilizing relation \eqref{eq80} into the particle's mass inequality yields:
\begin{equation}
\label{eq82}
m < \frac{1}{\sqrt{\phi^{2} (r_{m}) \alpha(r_{m})}}\left( E - \frac{q Q}{r_m} \right).
\end{equation}

Therefore, to overcharge a LyST black hole with $\left\{Q, M\right\}$ to the point that a naked singularity emerges it is essential to: choose $q$ which satisfies the equation \eqref{eq79} with the default $\left\{Q, M\right\}$, the particle's energy inequalities \eqref{eq77} and \eqref{eq78} must then be simultaneously satisfied and it is further required that $m$ obeys to \eqref{eq82} by using the previously defined $q$ and $E$. These steps assure that a charged test particle falls past the exterior horizon and overcharge the black hole to a final state $\left\{ Q_\text{f}, M_\text{f} \right\}$, such that:
\begin{equation}
\label{eq83}
\begin{cases}
Q_\text{f} = Q + q \\
4M_\text{f}^{\pm} = \frac{1}{r_{L}} \left( \sqrt{{r_{L}}^{2} + 4M r_{L}} \pm 2E \right)^{2} - r_{L} 
\end{cases},
\end{equation}
in which the final mass $M_\text{f}^{+}$ corresponds to the case $r_{L} > 0$ and $M_\text{f}^{-}$ to $r_{L} < -4M$.

\subsection{Eternal black hole} 
\label{subsubsec5.2}

A new interesting possibility within Lyra black holes is the existence of eternal overcharged solutions, a physical scenario which is impossible in the Reissner-Nodström spacetime. Nevertheless, there is no physical plausibility of some cases such as $Q \rightarrow Q_{\pm}$, since the exterior horizon goes to positive infinity for these charge values (see Figure \ref{fig: event and cauchy horizons}). As mentioned before, the final charge and mass of the black hole must obey new conditions such that:
\begin{equation}
\label{eq84}
\begin{cases}
E < Q - {}_{+}Q^{\text{x}}_{+} + q, \ \text{for} \ r_{L} > 0 \\
E < Q - {}_{-}Q^{\text{x}}_{-} + q, \ \text{for} \ r_{L} < -4M
\end{cases}.
\end{equation}
By combining these relations with \eqref{eq77} it is easy to obtain an inequality for the charge $q$:
\begin{equation}
\label{eq85}
q > \left(1 - \frac{Q}{r_+} \right)^{-1} \left(Q^{\text{x}} - Q\right),
\end{equation}
in which $Q^{\text{x}} = {}_{+}Q^{\text{x}}_{+}$ when $r_{L} > 0$ and $Q^{\text{x}} = {}_{-}Q^{\text{x}}_{-}$ for $r_{L} < -4M$. Moreover, as mentioned before, it is further considered that the $r_{L}$ value is close enough to $0$ or $-4M$ as to satisfy ${}_{+}Q^{\text{x}}_{+} - {}_{-}Q^{\text{x}}_{-} < q$ for $r_{L} > 0$ or ${}_{-}Q^{\text{x}}_{-} - {}_{-}Q^{\text{x}}_{+} < q$ when $r_{L} < -4M$, which results in:
\begin{equation}
\label{eq86}
\left\{ \begin{array}{ll}
     r_{L} < q, &\text{for}  ~r_{L} > 0 \\
\sqrt{r_{L}^{2} + 4Mr_{L}} < q, & \text{for} ~ r_{L} < -4M
\end{array}, \right.
\end{equation}
given that $0 < q \ll Q < {}_{\pm}Q^{\text{x}}_{\pm}$.

As a consequence, to overcharge a black hole with $\left\{Q, M\right\}$ as to surpass the charge intervals in which a naked singularity is possible, it is necessary that the particle's charge satisfies the inequality \eqref{eq85}, its energy must be within the bounds defined in \eqref{eq84} and \eqref{eq77}, such that $m$ obeys \eqref{eq82} with the previously defined $E$ and $q$. If these steps are completed, such that the $r_{L}$ value enables the inequalities in \eqref{eq86}, a charged particle $q$ with energy $E$ and mass $m$, for neglible backreaction effects, is able to fall into the black hole and overcharge it so that no further increase in its charge is capable of dissolving its horizons. This eternal black hole has then a charge $Q_\text{f}$ and a mass $M_\text{f}$ defined by the expressions in \eqref{eq83}.

\section{Final remarks}
\label{sec6}
We have constructed a generalization of Maxwell's electrodynamics for the Lyra geometry, which enabled us to find an analytical solution for the metric of a charged spherically symmetric black hole in the Lyra Scalar-Tensor Theory \cite{Cuzinatto_2021}. It was shown that one of the Lyra scale function effects is to compactify spacetime, such that for an infinite Lyra scale radius the Reissner-Nodström solution is reobtained. In the absence of electric charges, our solution proved to be consistent with the one obtained in \cite{Cuzinatto_2021}. Moreover, the Kretschmann scalar for Lyra manifolds reveals that a physical singularity is present at the origin. An analysis involving null and timelike geodesics additionally indicates that the $r_{-}$ and $r_{+}$ coordinate singularities are, respectively, a Cauchy and an apparent horizon.

It was further shown that, due to the presence of the Lyra scale radius, an effective cosmological constant arises from the black hole mass and charge when we consider the metric components alone. This quantity main effect is to alter the relation between the horizons and the mass and charge of the black hole, such that a black hole with a charge value greater than its geometrical mass is conceivable to exist within Lyra manifolds. It is also revealed the presence of six extremal charges, which divide the black hole horizons anatomy into different phenomenological cases: a singularity enclosed by two horizons, a naked singularity and two unique types of extremal black holes defined by relations \eqref{eq62} and \eqref{eq64}. 

These different black hole varieties are separated into three classes: $r_{L} > 0$ or $r_{L} < -4M$, $-4M \leq r_{L} < -2M$ and $-2M \leq r_{L} < 0$. All types of black holes mentioned before are possible in the first class, such that the horizons configurations of the positive charge interval limited by the second extremal charge are similar to the ones in charged black holes of General Relativity. A third extremal charge produces an unphysical scenario in which the exterior horizon diverges, such that for larger charge values it simply vanishes, so what was initially the inner horizon evolves into a lonely apparent horizon.  

The second class is similar to the first one, although a naked singularity is impossible to exist in this case. As for the third class, it has only one horizon, that is to be regarded as unphysical, since in the absence of charges no black hole would exist. This is further highlighted by the fact that for large charge values, a black hole in this class would span all of the tridimensional space. Therefore, values of the Lyra scale radius in the third class, i.e. $-2M \leq r_{L} < 0$, are to be disregarded.

Furthermore, we have studied the overcharging process of LyST charged black holes. It was considered radially infalling test charged particles as to overcharge our solution. A detailed analysis was constructed as to produce a naked singularity for the cases in which $r_{L} > 0$ or $r_{L} < -4M$. A similar process was constructed for a new possibility that emerges within the Lyra geometry, that is, an eternal overcharged black hole. It was further noted that, unlike what happens at the Reissner-Nodström metric, in the LyST theory it is possible to overcharge an extremal black hole. This analysis serves as a preliminary study to further consider backreaction effects as to verify if the weak cosmic censorship still holds in the Lyra Scalar-Tensor Theory. In addition, it is also important to consider an external magnetic field \cite{Gibbons2013}, since the overcharging process can be prevented in some cases \cite{Shaymatov2023}.

The study of LyST charged black holes and its different possible horizons configurations is very important, it is a preliminary analysis of how rotating black holes behave in the Lyra geometry. This is expected because the relation between the horizons and the mass and charge in the Reissner-Nodström solution is similar to what happens in the Kerr metric but with the presence of the angular momentum instead of the charge, and given that the solutions of General Relativity are always subcases of the LyST ones. Therefore, for example, by observing black hole shadows, it would be possible to measure deviations from Einstein's theory through the parameter $u_{L} = 1/r_{L}$. Thus, if General Relativity is compatible with observational data, it is expected that $u_{L} \sim 0$. 

Additionally, we are currently probing a more fundamental relation between the scalar-tensor theories obtained in the Lyra geometry, scale transformations of the metric components and the Weyl Integrable Spacetimes. This relation motivated us to pursue an unique Lyra theory with interesting cosmological applications. We are further studying rotating black holes in Lyra manifolds by considering a scale function that possess only a radial dependence. Electromagnetic phenomena arising from the Maxwell-Lyra equations are also been investigated. All these results will soon be published.

\section*{Acknowledgments}

We are grateful to professor José Abdalla Helayël-Neto and Isaque Porto de Freitas for the insightful discussions that greatly enriched this work. We are also thankful to the valuable assistance of professor Martín Makler and of our colleagues Henrique Santos Lima and Pedro Paraguassú. This project was only made possible thanks to the financial support provided by CNPq and FAPERJ.

\bibliography{apssamp}

\begin{thebibliography}{43}%
\makeatletter
\providecommand \@ifxundefined [1]{%
 \@ifx{#1\undefined}
}%
\providecommand \@ifnum [1]{%
 \ifnum #1\expandafter \@firstoftwo
 \else \expandafter \@secondoftwo
 \fi
}%
\providecommand \@ifx [1]{%
 \ifx #1\expandafter \@firstoftwo
 \else \expandafter \@secondoftwo
 \fi
}%
\providecommand \natexlab [1]{#1}%
\providecommand \enquote  [1]{``#1''}%
\providecommand \bibnamefont  [1]{#1}%
\providecommand \bibfnamefont [1]{#1}%
\providecommand \citenamefont [1]{#1}%
\providecommand \href@noop [0]{\@secondoftwo}%
\providecommand \href [0]{\begingroup \@sanitize@url \@href}%
\providecommand \@href[1]{\@@startlink{#1}\@@href}%
\providecommand \@@href[1]{\endgroup#1\@@endlink}%
\providecommand \@sanitize@url [0]{\catcode `\\12\catcode `\$12\catcode `\&12\catcode `\#12\catcode `\^12\catcode `\_12\catcode `\%12\relax}%
\providecommand \@@startlink[1]{}%
\providecommand \@@endlink[0]{}%
\providecommand \url  [0]{\begingroup\@sanitize@url \@url }%
\providecommand \@url [1]{\endgroup\@href {#1}{\urlprefix }}%
\providecommand \urlprefix  [0]{URL }%
\providecommand \Eprint [0]{\href }%
\providecommand \doibase [0]{https://doi.org/}%
\providecommand \selectlanguage [0]{\@gobble}%
\providecommand \bibinfo  [0]{\@secondoftwo}%
\providecommand \bibfield  [0]{\@secondoftwo}%
\providecommand \translation [1]{[#1]}%
\providecommand \BibitemOpen [0]{}%
\providecommand \bibitemStop [0]{}%
\providecommand \bibitemNoStop [0]{.\EOS\space}%
\providecommand \EOS [0]{\spacefactor3000\relax}%
\providecommand \BibitemShut  [1]{\csname bibitem#1\endcsname}%
\let\auto@bib@innerbib\@empty
\bibitem [{\citenamefont {{Abbott \textit{et al.}}}(2019{\natexlab{a}})}]{Abbott_2019a}%
  \BibitemOpen
  \bibfield  {author} {\bibinfo {author} {\bibfnamefont {B.~P.}\ \bibnamefont {{Abbott \textit{et al.}}}} (\bibinfo {collaboration} {Virgo, LIGO Scientific Collaboration}),\ }\bibfield  {title} {\bibinfo {title} {``{{GWTC}-1: A Gravitational-Wave Transient Catalog of Compact Binary Mergers Observed by {LIGO} and Virgo during the First and Second Observing Runs}''},\ }\bibfield  {journal} {\bibinfo  {journal} {Phys. Rev. X}\ }\textbf {\bibinfo {volume} {9}},\ \href {https://doi.org/10.1103/physrevx.9.031040} {10.1103/physrevx.9.031040} (\bibinfo {year} {2019}{\natexlab{a}}),\ \Eprint {https://arxiv.org/abs/arXiv:1811.12907 [astro-ph.HE]} {arXiv:1811.12907 [astro-ph.HE]} \BibitemShut {NoStop}%
\bibitem [{\citenamefont {{Abbott \textit{et al.}}}(2019{\natexlab{b}})}]{Abbott_2019b}%
  \BibitemOpen
  \bibfield  {author} {\bibinfo {author} {\bibfnamefont {B.~P.}\ \bibnamefont {{Abbott \textit{et al.}}}} (\bibinfo {collaboration} {Virgo, LIGO Scientific Collaboration}),\ }\bibfield  {title} {\bibinfo {title} {``{Binary Black Hole Population Properties Inferred from the First and Second Observing Runs of Advanced {LIGO} and Advanced Virgo}''},\ }\href {https://doi.org/10.3847/2041-8213/ab3800} {\bibfield  {journal} {\bibinfo  {journal} {Astrophys. J.}\ }\textbf {\bibinfo {volume} {882}},\ \bibinfo {pages} {L24} (\bibinfo {year} {2019}{\natexlab{b}})},\ \Eprint {https://arxiv.org/abs/arXiv:1811.12940 [astro-ph.HE]} {arXiv:1811.12940 [astro-ph.HE]} \BibitemShut {NoStop}%
\bibitem [{\citenamefont {{Akiyama \textit{et al.}}}(2019{\natexlab{a}})}]{Akiyama_2019a}%
  \BibitemOpen
  \bibfield  {author} {\bibinfo {author} {\bibfnamefont {K.}~\bibnamefont {{Akiyama \textit{et al.}}}} (\bibinfo {collaboration} {The Event Horizon Telescope Collaboration}),\ }\bibfield  {title} {\bibinfo {title} {``{First M87 Event Horizon Telescope Results. I. The Shadow of the Supermassive Black Hole}''},\ }\href {https://doi.org/10.3847/2041-8213/ab0ec7} {\bibfield  {journal} {\bibinfo  {journal} {Astrophys. J. Lett.}\ }\textbf {\bibinfo {volume} {875}},\ \bibinfo {pages} {L1} (\bibinfo {year} {2019}{\natexlab{a}})},\ \Eprint {https://arxiv.org/abs/arXiv:1906.11238 [astro-ph.GA]} {arXiv:1906.11238 [astro-ph.GA]} \BibitemShut {NoStop}%
\bibitem [{\citenamefont {{Akiyama \textit{et al.}}}(2019{\natexlab{b}})}]{Akiyama_2019b}%
  \BibitemOpen
  \bibfield  {author} {\bibinfo {author} {\bibfnamefont {K.}~\bibnamefont {{Akiyama \textit{et al.}}}} (\bibinfo {collaboration} {The Event Horizon Telescope Collaboration}),\ }\bibfield  {title} {\bibinfo {title} {``{First M87 Event Horizon Telescope Results. II. Array and Instrumentation}''},\ }\href {https://doi.org/10.3847/2041-8213/ab0c96} {\bibfield  {journal} {\bibinfo  {journal} {Astrophys. J. Lett.}\ }\textbf {\bibinfo {volume} {875}},\ \bibinfo {pages} {L2} (\bibinfo {year} {2019}{\natexlab{b}})},\ \Eprint {https://arxiv.org/abs/arXiv:1906.11239 [astro-ph.IM]} {arXiv:1906.11239 [astro-ph.IM]} \BibitemShut {NoStop}%
\bibitem [{\citenamefont {{Akiyama \textit{et al.}}}(2019{\natexlab{c}})}]{Akiyama_2019c}%
  \BibitemOpen
  \bibfield  {author} {\bibinfo {author} {\bibfnamefont {K.}~\bibnamefont {{Akiyama \textit{et al.}}}} (\bibinfo {collaboration} {The Event Horizon Telescope Collaboration}),\ }\bibfield  {title} {\bibinfo {title} {``{First M87 Event Horizon Telescope Results. III. Data Processing and Calibration}''},\ }\href {https://doi.org/10.3847/2041-8213/ab0c57} {\bibfield  {journal} {\bibinfo  {journal} {Astrophys. J. Lett.}\ }\textbf {\bibinfo {volume} {875}},\ \bibinfo {pages} {L3} (\bibinfo {year} {2019}{\natexlab{c}})},\ \Eprint {https://arxiv.org/abs/arXiv:1906.11240 [astro-ph.GA]} {arXiv:1906.11240 [astro-ph.GA]} \BibitemShut {NoStop}%
\bibitem [{\citenamefont {{Akiyama \textit{et al.}}}(2019{\natexlab{d}})}]{Akiyama_2019d}%
  \BibitemOpen
  \bibfield  {author} {\bibinfo {author} {\bibfnamefont {K.}~\bibnamefont {{Akiyama \textit{et al.}}}} (\bibinfo {collaboration} {The Event Horizon Telescope Collaboration}),\ }\bibfield  {title} {\bibinfo {title} {``{First M87 Event Horizon Telescope Results. IV. Imaging the Central Supermassive Black Hole}''},\ }\href {https://doi.org/10.3847/2041-8213/ab0e85} {\bibfield  {journal} {\bibinfo  {journal} {Astrophys. J. Lett.}\ }\textbf {\bibinfo {volume} {875}},\ \bibinfo {pages} {L4} (\bibinfo {year} {2019}{\natexlab{d}})},\ \Eprint {https://arxiv.org/abs/arXiv:1906.11241 [astro-ph.GA]} {arXiv:1906.11241 [astro-ph.GA]} \BibitemShut {NoStop}%
\bibitem [{\citenamefont {{Akiyama \textit{et al.}}}(2019{\natexlab{e}})}]{Akiyama_2019e}%
  \BibitemOpen
  \bibfield  {author} {\bibinfo {author} {\bibfnamefont {K.}~\bibnamefont {{Akiyama \textit{et al.}}}} (\bibinfo {collaboration} {The Event Horizon Telescope Collaboration}),\ }\bibfield  {title} {\bibinfo {title} {``{First M87 Event Horizon Telescope Results. V. Physical Origin of the Asymmetric Ring}''},\ }\href {https://doi.org/10.3847/2041-8213/ab0f43} {\bibfield  {journal} {\bibinfo  {journal} {Astrophys. J. Lett.}\ }\textbf {\bibinfo {volume} {875}},\ \bibinfo {pages} {L5} (\bibinfo {year} {2019}{\natexlab{e}})},\ \Eprint {https://arxiv.org/abs/arXiv:1906.11242 [astro-ph.GA]} {arXiv:1906.11242 [astro-ph.GA]} \BibitemShut {NoStop}%
\bibitem [{\citenamefont {{Akiyama \textit{et al.}}}(2019{\natexlab{f}})}]{Akiyama_2019f}%
  \BibitemOpen
  \bibfield  {author} {\bibinfo {author} {\bibfnamefont {K.}~\bibnamefont {{Akiyama \textit{et al.}}}} (\bibinfo {collaboration} {The Event Horizon Telescope Collaboration}),\ }\bibfield  {title} {\bibinfo {title} {``{First M87 Event Horizon Telescope Results. VI. The Shadow and Mass of the Central Black Hole}''},\ }\href {https://doi.org/10.3847/2041-8213/ab1141} {\bibfield  {journal} {\bibinfo  {journal} {Astrophys. J. Lett.}\ }\textbf {\bibinfo {volume} {875}},\ \bibinfo {pages} {L6} (\bibinfo {year} {2019}{\natexlab{f}})},\ \Eprint {https://arxiv.org/abs/arXiv:1906.11243 [astro-ph.GA]} {arXiv:1906.11243 [astro-ph.GA]} \BibitemShut {NoStop}%
\bibitem [{\citenamefont {{Akiyama \textit{et al.}}}(2021{\natexlab{a}})}]{Akiyama_2021a}%
  \BibitemOpen
  \bibfield  {author} {\bibinfo {author} {\bibfnamefont {K.}~\bibnamefont {{Akiyama \textit{et al.}}}} (\bibinfo {collaboration} {The Event Horizon Telescope Collaboration}),\ }\bibfield  {title} {\bibinfo {title} {``{First M87 Event Horizon Telescope Results. VII. Polarization of the Ring}''},\ }\href {https://doi.org/10.3847/2041-8213/abe71d} {\bibfield  {journal} {\bibinfo  {journal} {Astrophys. J. Lett.}\ }\textbf {\bibinfo {volume} {910}},\ \bibinfo {pages} {L12} (\bibinfo {year} {2021}{\natexlab{a}})},\ \Eprint {https://arxiv.org/abs/arXiv:2105.01169 [astro-ph.HE]} {arXiv:2105.01169 [astro-ph.HE]} \BibitemShut {NoStop}%
\bibitem [{\citenamefont {{Akiyama \textit{et al.}}}(2021{\natexlab{b}})}]{Akiyama_2021b}%
  \BibitemOpen
  \bibfield  {author} {\bibinfo {author} {\bibfnamefont {K.}~\bibnamefont {{Akiyama \textit{et al.}}}} (\bibinfo {collaboration} {The Event Horizon Telescope Collaboration}),\ }\bibfield  {title} {\bibinfo {title} {``{First M87 Event Horizon Telescope Results. VIII. Magnetic Field Structure near The Event Horizon}''},\ }\href {https://doi.org/10.3847/2041-8213/abe4de} {\bibfield  {journal} {\bibinfo  {journal} {Astrophys. J. Lett.}\ }\textbf {\bibinfo {volume} {910}},\ \bibinfo {pages} {L13} (\bibinfo {year} {2021}{\natexlab{b}})},\ \Eprint {https://arxiv.org/abs/arXiv:2105.01173 [astro-ph.HE]} {arXiv:2105.01173 [astro-ph.HE]} \BibitemShut {NoStop}%
\bibitem [{\citenamefont {Medeiros}\ \emph {et~al.}(2023)\citenamefont {Medeiros}, \citenamefont {Psaltis}, \citenamefont {Lauer},\ and\ \citenamefont {Özel}}]{Medeiros_2023}%
  \BibitemOpen
  \bibfield  {author} {\bibinfo {author} {\bibfnamefont {L.}~\bibnamefont {Medeiros}}, \bibinfo {author} {\bibfnamefont {D.}~\bibnamefont {Psaltis}}, \bibinfo {author} {\bibfnamefont {T.~R.}\ \bibnamefont {Lauer}},\ and\ \bibinfo {author} {\bibfnamefont {F.}~\bibnamefont {Özel}},\ }\bibfield  {title} {\bibinfo {title} {``{The Image of the M87 Black Hole Reconstructed with PRIMO}''},\ }\href {https://doi.org/10.3847/2041-8213/acc32d} {\bibfield  {journal} {\bibinfo  {journal} {Astrophys. J. Lett.}\ }\textbf {\bibinfo {volume} {947}},\ \bibinfo {pages} {L7} (\bibinfo {year} {2023})},\ \Eprint {https://arxiv.org/abs/arXiv:2304.06079 [astro-ph.HE]} {arXiv:2304.06079 [astro-ph.HE]} \BibitemShut {NoStop}%
\bibitem [{\citenamefont {{Xu \textit{et al.}}}(2023)}]{Xu_2023}%
  \BibitemOpen
  \bibfield  {author} {\bibinfo {author} {\bibfnamefont {H.}~\bibnamefont {{Xu \textit{et al.}}}} (\bibinfo {collaboration} {The Chinese Pulsar Timing Array Collaboration}),\ }\bibfield  {title} {\bibinfo {title} {``{Searching for the Nano-Hertz Stochastic Gravitational Wave Background with the Chinese Pulsar Timing Array Data Release I}''},\ }\href {https://doi.org/10.1088/1674-4527/acdfa5} {\bibfield  {journal} {\bibinfo  {journal} {Res. Astron. Astrophys.}\ }\textbf {\bibinfo {volume} {23}},\ \bibinfo {pages} {075024} (\bibinfo {year} {2023})},\ \Eprint {https://arxiv.org/abs/arXiv:2306.16216 [astro-ph.HE]} {arXiv:2306.16216 [astro-ph.HE]} \BibitemShut {NoStop}%
\bibitem [{\citenamefont {{Antoniadis \textit{et al.}}}(2023)}]{antoniadis2023second}%
  \BibitemOpen
  \bibfield  {author} {\bibinfo {author} {\bibfnamefont {J.}~\bibnamefont {{Antoniadis \textit{et al.}}}} (\bibinfo {collaboration} {EPTA Collaboration and InPTA Collaboration}),\ }\bibfield  {title} {\bibinfo {title} {``{The second data release from the European Pulsar Timing Array - III. Search for gravitational wave signals}''},\ }\href {https://doi.org/10.1051/0004-6361/202346844} {\bibfield  {journal} {\bibinfo  {journal} {Astron. Astrophys.}\ }\textbf {\bibinfo {volume} {678}},\ \bibinfo {pages} {A50} (\bibinfo {year} {2023})}\BibitemShut {NoStop}%
\bibitem [{\citenamefont {{Reardon \textit{et al.}}}(2023)}]{Reardon_2023}%
  \BibitemOpen
  \bibfield  {author} {\bibinfo {author} {\bibfnamefont {D.~J.}\ \bibnamefont {{Reardon \textit{et al.}}}} (\bibinfo {collaboration} {The International Pulsar Timing Array Collaboration}),\ }\bibfield  {title} {\bibinfo {title} {``{Search for an Isotropic Gravitational-wave Background with the Parkes Pulsar Timing Array}''},\ }\href {https://doi.org/10.3847/2041-8213/acdd02} {\bibfield  {journal} {\bibinfo  {journal} {Astrophys. J. Lett.}\ }\textbf {\bibinfo {volume} {951}},\ \bibinfo {pages} {L6} (\bibinfo {year} {2023})},\ \Eprint {https://arxiv.org/abs/arXiv:2306.16215 [astro-ph.HE]} {arXiv:2306.16215 [astro-ph.HE]} \BibitemShut {NoStop}%
\bibitem [{\citenamefont {{Agazie \textit{et al.}}}(2023{\natexlab{a}})}]{Agazie_2023a}%
  \BibitemOpen
  \bibfield  {author} {\bibinfo {author} {\bibfnamefont {G.}~\bibnamefont {{Agazie \textit{et al.}}}} (\bibinfo {collaboration} {The NANOGrav Collaboration}),\ }\bibfield  {title} {\bibinfo {title} {``{The NANOGrav 15 yr Data Set: Bayesian Limits on Gravitational Waves from Individual Supermassive Black Hole Binaries}''},\ }\href {https://doi.org/10.3847/2041-8213/ace18a} {\bibfield  {journal} {\bibinfo  {journal} {Astrophys. J. Lett.}\ }\textbf {\bibinfo {volume} {951}},\ \bibinfo {pages} {L50} (\bibinfo {year} {2023}{\natexlab{a}})},\ \Eprint {https://arxiv.org/abs/arXiv:2306.16213 [astro-ph.HE]} {arXiv:2306.16213 [astro-ph.HE]} \BibitemShut {NoStop}%
\bibitem [{\citenamefont {{Agazie \textit{et al.}}}(2023{\natexlab{b}})}]{Agazie_2023b}%
  \BibitemOpen
  \bibfield  {author} {\bibinfo {author} {\bibfnamefont {G.}~\bibnamefont {{Agazie \textit{et al.}}}} (\bibinfo {collaboration} {The NANOGrav Collaboration}),\ }\bibfield  {title} {\bibinfo {title} {``{The NANOGrav 15 yr Data Set: Bayesian Limits on Gravitational Waves from Individual Supermassive Black Hole Binaries}''},\ }\href {https://doi.org/10.3847/2041-8213/ace18a} {\bibfield  {journal} {\bibinfo  {journal} {Astrophys. J. Lett.}\ }\textbf {\bibinfo {volume} {951}},\ \bibinfo {pages} {L50} (\bibinfo {year} {2023}{\natexlab{b}})},\ \Eprint {https://arxiv.org/abs/arXiv:2306.16222 [astro-ph.HE]} {arXiv:2306.16222 [astro-ph.HE]} \BibitemShut {NoStop}%
\bibitem [{\citenamefont {Penrose}(1965)}]{Penrose_1965}%
  \BibitemOpen
  \bibfield  {author} {\bibinfo {author} {\bibfnamefont {R.}~\bibnamefont {Penrose}},\ }\bibfield  {title} {\bibinfo {title} {``{Gravitational Collapse and Space-Time Singularities}''},\ }\href {https://doi.org/10.1103/PhysRevLett.14.57} {\bibfield  {journal} {\bibinfo  {journal} {Phys. Rev. Lett.}\ }\textbf {\bibinfo {volume} {14}},\ \bibinfo {pages} {57} (\bibinfo {year} {1965})}\BibitemShut {NoStop}%
\bibitem [{\citenamefont {Penrose}(1973)}]{Penrose_1973}%
  \BibitemOpen
  \bibfield  {author} {\bibinfo {author} {\bibfnamefont {R.}~\bibnamefont {Penrose}},\ }\bibfield  {title} {\bibinfo {title} {``{Naked Singularities}''},\ }\href {https://doi.org/https://doi.org/10.1111/j.1749-6632.1973.tb41447.x} {\bibfield  {journal} {\bibinfo  {journal} {Ann. N. Y. Acad. Sci.}\ }\textbf {\bibinfo {volume} {224}},\ \bibinfo {pages} {125} (\bibinfo {year} {1973})}\BibitemShut {NoStop}%
\bibitem [{\citenamefont {Penrose}\ \emph {et~al.}(1993)\citenamefont {Penrose}, \citenamefont {Sorkin},\ and\ \citenamefont {Woolgar}}]{Penrose_1993}%
  \BibitemOpen
  \bibfield  {author} {\bibinfo {author} {\bibfnamefont {R.}~\bibnamefont {Penrose}}, \bibinfo {author} {\bibfnamefont {R.~D.}\ \bibnamefont {Sorkin}},\ and\ \bibinfo {author} {\bibfnamefont {E.}~\bibnamefont {Woolgar}},\ }\href@noop {} {\bibinfo {title} {``{A Positive mass theorem based on the focusing and retardation of null geodesics}''}},\ \bibinfo {howpublished} {e-print arXiv:gr-qc/9301015} (\bibinfo {year} {1993})\BibitemShut {NoStop}%
\bibitem [{\citenamefont {Penrose}(1999)}]{Penrose_1999}%
  \BibitemOpen
  \bibfield  {author} {\bibinfo {author} {\bibfnamefont {R.}~\bibnamefont {Penrose}},\ }\bibfield  {title} {\bibinfo {title} {``{The question of cosmic censorship}''},\ }\href {https://doi.org/https://doi.org/10.1007/BF02702355} {\bibfield  {journal} {\bibinfo  {journal} {J. Astrophys. Astron.}\ }\textbf {\bibinfo {volume} {20}},\ \bibinfo {pages} {233} (\bibinfo {year} {1999})}\BibitemShut {NoStop}%
\bibitem [{\citenamefont {Kerr}(2023)}]{Kerr2023rpn}%
  \BibitemOpen
  \bibfield  {author} {\bibinfo {author} {\bibfnamefont {R.~P.}\ \bibnamefont {Kerr}},\ }\href {https://arxiv.org/abs/2312.00841} {\bibinfo {title} {``{Do Black Holes have Singularities?}''}},\ \bibinfo {howpublished} {e-print arXiv:2312.00841[gr-qc]} (\bibinfo {year} {2023})\BibitemShut {NoStop}%
\bibitem [{\citenamefont {Hawking}(2005)}]{Hawking_2005}%
  \BibitemOpen
  \bibfield  {author} {\bibinfo {author} {\bibfnamefont {S.~W.}\ \bibnamefont {Hawking}},\ }\bibfield  {title} {\bibinfo {title} {``{Information loss in black holes}''},\ }\bibfield  {journal} {\bibinfo  {journal} {Phys. Rev. D}\ }\textbf {\bibinfo {volume} {72}},\ \href {https://doi.org/10.1103/physrevd.72.084013} {10.1103/physrevd.72.084013} (\bibinfo {year} {2005}),\ \Eprint {https://arxiv.org/abs/arXiv:hep-th/0507171} {arXiv:hep-th/0507171} \BibitemShut {NoStop}%
\bibitem [{\citenamefont {Frampton}\ \emph {et~al.}(2010)\citenamefont {Frampton}, \citenamefont {Kawasaki}, \citenamefont {Takahashi},\ and\ \citenamefont {Yanagida}}]{Frampton_2010}%
  \BibitemOpen
  \bibfield  {author} {\bibinfo {author} {\bibfnamefont {P.~H.}\ \bibnamefont {Frampton}}, \bibinfo {author} {\bibfnamefont {M.}~\bibnamefont {Kawasaki}}, \bibinfo {author} {\bibfnamefont {F.}~\bibnamefont {Takahashi}},\ and\ \bibinfo {author} {\bibfnamefont {T.~T.}\ \bibnamefont {Yanagida}},\ }\bibfield  {title} {\bibinfo {title} {``{Primordial black holes as all dark matter}''},\ }\href {https://doi.org/10.1088/1475-7516/2010/04/023} {\bibfield  {journal} {\bibinfo  {journal} {J. Cosmol. Astropart. Phys.}\ }\textbf {\bibinfo {volume} {2010}}\bibfield  {number} {\bibinfo  {number} { (04)},\ \bibinfo {pages} {023}},\ }\Eprint {https://arxiv.org/abs/arXiv:1001.2308 [hep-ph]} {arXiv:1001.2308 [hep-ph]} \BibitemShut {NoStop}%
\bibitem [{\citenamefont {Cappelluti}\ \emph {et~al.}(2022)\citenamefont {Cappelluti}, \citenamefont {Hasinger},\ and\ \citenamefont {Natarajan}}]{Cappelluti_2022}%
  \BibitemOpen
  \bibfield  {author} {\bibinfo {author} {\bibfnamefont {N.}~\bibnamefont {Cappelluti}}, \bibinfo {author} {\bibfnamefont {G.}~\bibnamefont {Hasinger}},\ and\ \bibinfo {author} {\bibfnamefont {P.}~\bibnamefont {Natarajan}},\ }\bibfield  {title} {\bibinfo {title} {``{Exploring the High-redshift PBH-$\Lambda$CDM Universe: Early Black Hole Seeding, the First Stars and Cosmic Radiation Backgrounds}''},\ }\href {https://doi.org/10.3847/1538-4357/ac332d} {\bibfield  {journal} {\bibinfo  {journal} {Astrophys. J.}\ }\textbf {\bibinfo {volume} {926}},\ \bibinfo {pages} {205} (\bibinfo {year} {2022})},\ \Eprint {https://arxiv.org/abs/arXiv:2109.08701 [astro-ph.CO]} {arXiv:2109.08701 [astro-ph.CO]} \BibitemShut {NoStop}%
\bibitem [{\citenamefont {Croker}\ and\ \citenamefont {Weiner}(2019)}]{Croker_2019}%
  \BibitemOpen
  \bibfield  {author} {\bibinfo {author} {\bibfnamefont {K.~S.}\ \bibnamefont {Croker}}\ and\ \bibinfo {author} {\bibfnamefont {J.~L.}\ \bibnamefont {Weiner}},\ }\bibfield  {title} {\bibinfo {title} {``{Implications of Symmetry and Pressure in Friedmann Cosmology. I. Formalism}''},\ }\href {https://doi.org/10.3847/1538-4357/ab32da} {\bibfield  {journal} {\bibinfo  {journal} {Astrophys. J.}\ }\textbf {\bibinfo {volume} {882}},\ \bibinfo {pages} {19} (\bibinfo {year} {2019})},\ \Eprint {https://arxiv.org/abs/arXiv:2107.06643 [gr-qc]} {arXiv:2107.06643 [gr-qc]} \BibitemShut {NoStop}%
\bibitem [{\citenamefont {{Farrah \textit{et al.}}}(2023)}]{Farrah_2023}%
  \BibitemOpen
  \bibfield  {author} {\bibinfo {author} {\bibfnamefont {D.}~\bibnamefont {{Farrah \textit{et al.}}}},\ }\bibfield  {title} {\bibinfo {title} {``{Observational Evidence for Cosmological Coupling of Black Holes and its Implications for an Astrophysical Source of Dark Energy}''},\ }\href {https://doi.org/10.3847/2041-8213/acb704} {\bibfield  {journal} {\bibinfo  {journal} {Astrophys. J. Lett.}\ }\textbf {\bibinfo {volume} {944}},\ \bibinfo {pages} {L31} (\bibinfo {year} {2023})},\ \Eprint {https://arxiv.org/abs/arXiv:2302.07878 [astro-ph.CO]} {arXiv:2302.07878 [astro-ph.CO]} \BibitemShut {NoStop}%
\bibitem [{\citenamefont {Lyra}(1951)}]{Lyra_1951}%
  \BibitemOpen
  \bibfield  {author} {\bibinfo {author} {\bibfnamefont {G.}~\bibnamefont {Lyra}},\ }\bibfield  {title} {\bibinfo {title} {``{Über eine Modifikation der Riemannschen Geometrie}''},\ }\href {https://doi.org/https://doi.org/10.1007/BF01175135} {\bibfield  {journal} {\bibinfo  {journal} {Math. Z.}\ }\textbf {\bibinfo {volume} {54}},\ \bibinfo {pages} {52} (\bibinfo {year} {1951})}\BibitemShut {NoStop}%
\bibitem [{\citenamefont {Weyl}(1952)}]{Weyl_1922}%
  \BibitemOpen
  \bibfield  {author} {\bibinfo {author} {\bibfnamefont {H.}~\bibnamefont {Weyl}},\ }\href@noop {} {\emph {\bibinfo {title} {``{Space, Time, Matter}''}}}\ (\bibinfo  {publisher} {{Dover Books on Advanced Mathematics}},\ \bibinfo {address} {Dover, New York},\ \bibinfo {year} {1952})\BibitemShut {NoStop}%
\bibitem [{\citenamefont {Ross}(1972)}]{Ross_1972}%
  \BibitemOpen
  \bibfield  {author} {\bibinfo {author} {\bibfnamefont {D.~K.}\ \bibnamefont {Ross}},\ }\bibfield  {title} {\bibinfo {title} {``{Scalar-Tensor Theory of Gravitation}''},\ }\href {https://doi.org/10.1103/PhysRevD.5.284} {\bibfield  {journal} {\bibinfo  {journal} {Phys. Rev. D}\ }\textbf {\bibinfo {volume} {5}},\ \bibinfo {pages} {284} (\bibinfo {year} {1972})}\BibitemShut {NoStop}%
\bibitem [{\citenamefont {Novello}\ and\ \citenamefont {Heintzmann}(1983)}]{Novello_1983}%
  \BibitemOpen
  \bibfield  {author} {\bibinfo {author} {\bibfnamefont {M.}~\bibnamefont {Novello}}\ and\ \bibinfo {author} {\bibfnamefont {H.}~\bibnamefont {Heintzmann}},\ }\bibfield  {title} {\bibinfo {title} {``{Weyl integrable space-time: A model of our cosmos?}''},\ }\href {https://doi.org/10.1016/0375-9601(83)90532-7} {\bibfield  {journal} {\bibinfo  {journal} {Phys. Lett. A}\ }\textbf {\bibinfo {volume} {98}},\ \bibinfo {pages} {10} (\bibinfo {year} {1983})}\BibitemShut {NoStop}%
\bibitem [{\citenamefont {Sen}(1957)}]{Sen_1957}%
  \BibitemOpen
  \bibfield  {author} {\bibinfo {author} {\bibfnamefont {D.~K.}\ \bibnamefont {Sen}},\ }\bibfield  {title} {\bibinfo {title} {``{A static cosmological model}''},\ }\href {https://doi.org/10.1007/BF01333146} {\bibfield  {journal} {\bibinfo  {journal} {Z. Phys.}\ }\textbf {\bibinfo {volume} {149}},\ \bibinfo {pages} {311} (\bibinfo {year} {1957})}\BibitemShut {NoStop}%
\bibitem [{\citenamefont {Sen}\ and\ \citenamefont {Dunn}(1971)}]{Sen_1971}%
  \BibitemOpen
  \bibfield  {author} {\bibinfo {author} {\bibfnamefont {D.~K.}\ \bibnamefont {Sen}}\ and\ \bibinfo {author} {\bibfnamefont {K.~A.}\ \bibnamefont {Dunn}},\ }\bibfield  {title} {\bibinfo {title} {``{A Scalar-Tensor Theory of Gravitation in a Modified Riemannian Manifold}''},\ }\href {https://doi.org/10.1063/1.1665623} {\bibfield  {journal} {\bibinfo  {journal} {J. Math. Phys.}\ }\textbf {\bibinfo {volume} {12}},\ \bibinfo {pages} {578} (\bibinfo {year} {1971})}\BibitemShut {NoStop}%
\bibitem [{\citenamefont {Jeavons}\ \emph {et~al.}(1975)\citenamefont {Jeavons}, \citenamefont {McIntosh},\ and\ \citenamefont {Sen}}]{Jeavons_1975}%
  \BibitemOpen
  \bibfield  {author} {\bibinfo {author} {\bibfnamefont {J.~S.}\ \bibnamefont {Jeavons}}, \bibinfo {author} {\bibfnamefont {C.~B.~G.}\ \bibnamefont {McIntosh}},\ and\ \bibinfo {author} {\bibfnamefont {D.~K.}\ \bibnamefont {Sen}},\ }\bibfield  {title} {\bibinfo {title} {``{A correction to the Sen and Dunn gravitational field equations}''},\ }\href {https://doi.org/10.1063/1.522544} {\bibfield  {journal} {\bibinfo  {journal} {J. Math. Phys.}\ }\textbf {\bibinfo {volume} {16}},\ \bibinfo {pages} {320} (\bibinfo {year} {1975})}\BibitemShut {NoStop}%
\bibitem [{\citenamefont {Cuzinatto}\ \emph {et~al.}(2021)\citenamefont {Cuzinatto}, \citenamefont {de~Morais},\ and\ \citenamefont {Pimentel}}]{Cuzinatto_2021}%
  \BibitemOpen
  \bibfield  {author} {\bibinfo {author} {\bibfnamefont {R.~R.}\ \bibnamefont {Cuzinatto}}, \bibinfo {author} {\bibfnamefont {E.~M.}\ \bibnamefont {de~Morais}},\ and\ \bibinfo {author} {\bibfnamefont {B.~M.}\ \bibnamefont {Pimentel}},\ }\bibfield  {title} {\bibinfo {title} {``{Lyra scalar-tensor theory: A scalar-tensor theory of gravity on Lyra manifold}''},\ }\bibfield  {journal} {\bibinfo  {journal} {Phys. Rev. D}\ }\textbf {\bibinfo {volume} {103}},\ \href {https://doi.org/10.1103/physrevd.103.124002} {10.1103/physrevd.103.124002} (\bibinfo {year} {2021})\BibitemShut {NoStop}%
\bibitem [{\citenamefont {Weyl}(1918)}]{Weyl_1918}%
  \BibitemOpen
  \bibfield  {author} {\bibinfo {author} {\bibfnamefont {H.}~\bibnamefont {Weyl}},\ }\bibfield  {title} {\bibinfo {title} {``{Gravitation und Elektrizit{\"a}t}''},\ }\href {https://doi.org/10.1007/978-3-663-19510-8_11} {\bibfield  {journal} {\bibinfo  {journal} {Sitzungsber. Preuss. Akad. Wiss.}\ ,\ \bibinfo {pages} {465}} (\bibinfo {year} {1918})}\BibitemShut {NoStop}%
\bibitem [{\citenamefont {Dirac}(1973)}]{Dirac_1973}%
  \BibitemOpen
  \bibfield  {author} {\bibinfo {author} {\bibfnamefont {P.~A.~M.}\ \bibnamefont {Dirac}},\ }\bibfield  {title} {\bibinfo {title} {``{Long Range Forces and Broken Symmetries}''},\ }\href {https://doi.org/10.1098/rspa.1973.0070} {\bibfield  {journal} {\bibinfo  {journal} {Proc. R. Soc. London Ser. A}\ }\textbf {\bibinfo {volume} {333}},\ \bibinfo {pages} {403} (\bibinfo {year} {1973})}\BibitemShut {NoStop}%
\bibitem [{\citenamefont {Maeder}(1978)}]{Maeder_1978}%
  \BibitemOpen
  \bibfield  {author} {\bibinfo {author} {\bibfnamefont {A.}~\bibnamefont {Maeder}},\ }\bibfield  {title} {\bibinfo {title} {``{Metrical connection in space-time, Newton's and Hubble's laws}''},\ }\href {https://ui.adsabs.harvard.edu/abs/1978A&A....65..337M} {\bibfield  {journal} {\bibinfo  {journal} {Astron. Astrophys.}\ }\textbf {\bibinfo {volume} {65}},\ \bibinfo {pages} {337} (\bibinfo {year} {1978})}\BibitemShut {NoStop}%
\bibitem [{\citenamefont {Casana}\ \emph {et~al.}(2006)\citenamefont {Casana}, \citenamefont {de~Melo},\ and\ \citenamefont {Pimentel}}]{Casana_2006}%
  \BibitemOpen
  \bibfield  {author} {\bibinfo {author} {\bibfnamefont {R.}~\bibnamefont {Casana}}, \bibinfo {author} {\bibfnamefont {C.~A.~M.}\ \bibnamefont {de~Melo}},\ and\ \bibinfo {author} {\bibfnamefont {B.~M.}\ \bibnamefont {Pimentel}},\ }\bibfield  {title} {\bibinfo {title} {``{Spinorial Field and Lyra Geometry}''},\ }\href {https://doi.org/10.1007/s10509-006-9048-5} {\bibfield  {journal} {\bibinfo  {journal} {Astrophys. Space Sc.}\ }\textbf {\bibinfo {volume} {305}},\ \bibinfo {pages} {125} (\bibinfo {year} {2006})}\BibitemShut {NoStop}%
\bibitem [{\citenamefont {Scheibe}(1952)}]{Scheibe_1952}%
  \BibitemOpen
  \bibfield  {author} {\bibinfo {author} {\bibfnamefont {E.}~\bibnamefont {Scheibe}},\ }\bibfield  {title} {\bibinfo {title} {``{Über einen verallgemeinerten affinen Zusammenhang}''},\ }\href {https://doi.org/10.1007/BF01192916} {\bibfield  {journal} {\bibinfo  {journal} {Math. Z.}\ }\textbf {\bibinfo {volume} {57}},\ \bibinfo {pages} {65} (\bibinfo {year} {1952})}\BibitemShut {NoStop}%
\bibitem [{\citenamefont {Romero}\ \emph {et~al.}(2019)\citenamefont {Romero}, \citenamefont {Lima},\ and\ \citenamefont {Sanomiya}}]{Romero_2019}%
  \BibitemOpen
  \bibfield  {author} {\bibinfo {author} {\bibfnamefont {C.}~\bibnamefont {Romero}}, \bibinfo {author} {\bibfnamefont {R.~G.}\ \bibnamefont {Lima}},\ and\ \bibinfo {author} {\bibfnamefont {T.~A.~T.}\ \bibnamefont {Sanomiya}},\ }\bibfield  {title} {\bibinfo {title} {``{One hundred years of Weyl’s (unfinished) unified field theory}''},\ }\href {https://doi.org/https://doi.org/10.1016/j.shpsb.2019.02.005} {\bibfield  {journal} {\bibinfo  {journal} {Stud. Hist. Philos. Sc. Part B: Mod. Phys.}\ }\textbf {\bibinfo {volume} {66}},\ \bibinfo {pages} {180} (\bibinfo {year} {2019})}\BibitemShut {NoStop}%
\bibitem [{\citenamefont {Hubeny}(1999)}]{Hubeny1999}%
  \BibitemOpen
  \bibfield  {author} {\bibinfo {author} {\bibfnamefont {V.~E.}\ \bibnamefont {Hubeny}},\ }\bibfield  {title} {\bibinfo {title} {``{Overcharging a black hole and cosmic censorship}''},\ }\href {https://doi.org/10.1103/PhysRevD.59.064013} {\bibfield  {journal} {\bibinfo  {journal} {Phys. Rev. D}\ }\textbf {\bibinfo {volume} {59}},\ \bibinfo {pages} {064013} (\bibinfo {year} {1999})}\BibitemShut {NoStop}%
\bibitem [{\citenamefont {Gibbons}\ \emph {et~al.}(2013)\citenamefont {Gibbons}, \citenamefont {Mujtaba},\ and\ \citenamefont {Pope}}]{Gibbons2013}%
  \BibitemOpen
  \bibfield  {author} {\bibinfo {author} {\bibfnamefont {G.}~\bibnamefont {Gibbons}}, \bibinfo {author} {\bibfnamefont {A.}~\bibnamefont {Mujtaba}},\ and\ \bibinfo {author} {\bibfnamefont {C.}~\bibnamefont {Pope}},\ }\bibfield  {title} {\bibinfo {title} {``{Ergoregions in magnetized black hole spacetimes}''},\ }\href {https://iopscience.iop.org/article/10.1088/0264-9381/30/12/125008} {\bibfield  {journal} {\bibinfo  {journal} {Class. Quantum Gravit.}\ }\textbf {\bibinfo {volume} {30}},\ \bibinfo {pages} {125008} (\bibinfo {year} {2013})}\BibitemShut {NoStop}%
\bibitem [{\citenamefont {Shaymatov}\ and\ \citenamefont {Ahmedov}(2023)}]{Shaymatov2023}%
  \BibitemOpen
  \bibfield  {author} {\bibinfo {author} {\bibfnamefont {S.}~\bibnamefont {Shaymatov}}\ and\ \bibinfo {author} {\bibfnamefont {B.}~\bibnamefont {Ahmedov}},\ }\bibfield  {title} {\bibinfo {title} {``{Overcharging process around a magnetized black hole: can the backreaction effect of magnetic field restore cosmic censorship conjecture?}''},\ }\bibfield  {journal} {\bibinfo  {journal} {Gen. Relativ. Gravit.}\ }\textbf {\bibinfo {volume} {55}},\ \href {https://doi.org/10.1007/s10714-023-03082-y} {10.1007/s10714-023-03082-y} (\bibinfo {year} {2023})\BibitemShut {NoStop}%
\end{thebibliography}%

\end{document}